\documentclass[preprint,aps,prc,showpacs,nofootinbib]{revtex4-1}
\usepackage{amsmath}
\usepackage{amssymb}
\usepackage{graphics}
\usepackage{epsfig}

\usepackage{subfigure}

\begin{document}
\title{Effects of excited electron and contact $ee\gamma\gamma $
interaction in $ e^+e^-\rightarrow \gamma\gamma $ reaction}

\author{G. I. Gakh}
\email{gakh@kipt.kharkov.ua}

\affiliation{\it National Science Centre, Kharkov Institute of
Physics and Technology, Akademicheskaya 1, and V. N. Karazin
Kharkov National University, Dept. of
Physics and Technology, 31 Kurchatov, 61108 Kharkov, Ukraine}
\author{M.I. Konchatnij}
 \email{konchatnij@kipt.kharkov.ua}

 \affiliation{\it National Science Centre, Kharkov Institute of
Physics and Technology, Akademicheskaya 1, and V. N. Karazin
Kharkov National University, Dept. of
Physics and Technology, 31 Kurchatov, 61108 Kharkov, Ukraine}
\author{N.P. Merenkov}
\email{merenkov@@kipt.kharkov.ua}
\affiliation{\it National Science Centre, Kharkov Institute of
Physics and Technology, Akademicheskaya 1, and V. N. Karazin
Kharkov
National University, Dept. of
Physics and Technology, 31 Kurchatov, 61108 Kharkov, Ukraine}

\author{A. G. Gakh}
\email{agakh@karazin.ua}

\affiliation{V. N. Karazin
Kharkov National University, Dept. of
Physics and Technology, 31 Kurchatov, 61108 Kharkov, Ukraine}



\begin{abstract}
The differential cross section and some polarization observables have been calculated for the $e^+\,e^-\to \gamma\,\gamma$
reaction taking into account the contribution of the heavy excited electron and the general form $e\,e\,\gamma\,\gamma$
contact interaction. The spin correlation coefficients are calculated for the case when both beams have arbitrary polarization.
The so called left-right asymmetry is estimated if only one beam is polarized. Numerical estimations are given for the
excited electron contribution to the differential cross section and spin correlation coefficients for various values of the
electron beam energy and excited electron mass.
\end{abstract}

\vspace{0.2cm}
\vspace{0.2cm}

\maketitle

\section{Introduction}
\hspace{0.7cm}

As it is known, there exist the experimental data which are not explained in the framework of the Standard Model (such as the presence
of a new form of matter, the so-called Dark Matter of the Universe, the Dark Energy and so on). The more detailed discussion of these
questions see the lectures \cite{BSM}. All these observations to the necessity of the requirement for the extension of the Standard
Model. There exist several approaches to do this.

One of the possible approaches is the so-called the composite models \cite{Comp}. Some fundamental questions which do not explained
by the Standard Model (SM)(such as the number of fermion families, the fermion masses, and the mixings) were supposed to solve in
the framework of these models. The existence of the excited states is the natural consequence of the composite models of quarks and
leptons. The increase of the number of quarks and leptons is often considered as the hint that these particles have substructure.
The most convincing proof that quarks and leptons have substructure would be the discovery of the excited states of ordinary
quarks and leptons (which are considered, in this case, as ground states): $l, l^*, l^{**},..., q, q^*, q^{**}, ..$ The most
simple hypothesis is to assign the fermions the same electroweak, colour and spin values as to its corresponding low--lying partners.
The excited states go to the ground states by means of generalized transition of the magnetic type \cite{L65} (emitting photon (for
the lepton case) or gluon (for the quark case)). In reality, for the excited states with small mass the radiative transition
is the main decay branching ratio. But when their masses approach to the $W$--boson mass the large fraction of the three--particle
final states appear in the decays of the excited states.

At present, there is no completely predictable model describing the substructure of the quarks and leptons. Therefore, the best thing,
that can be done for the search of the substructure effects, is to perform the necessary phenomenological analysis. The review of possible
effects  of the substructure of the quarks and leptons, which can be displayed in various reactions, is given in Ref. \cite{B91}. For
the investigation of possible lepton substructure QED is the ideal theory, because every deviation from QED $e^+e^-$ scattering in
differential or total cross section, could be interpreted as a deviation from the point - like structure of the electron or new physics.

The search of the excited charged (also neutral, at last time) fermions is systematically continued for more than 30 years but up to now
there is no success. For the case of heavy electron the mass less than 146 GeV \cite{A95} was already excluded experimentally. The case
of the excited quarks \cite{H85} is the direct generalization of the lepton case. From the theoretical point of view, the preferable
mechanism of the transition of the excited quark is the gluon emission $q^*\to qg$ (gluon has a large energy). However, it will
be very difficult to extract this effect from a rather large standard background of the three--jet events. Thus, from the experimental
point of view, the most preferable field for the search of the substructure effects is the lepton sector.

In the actual experimental investigation the differential cross section of the $e^+\,e^-\to \gamma\,\gamma$ reaction was used to study the
possibility of a deviation from QED using all experimental data generated from 1989 by different electron - positron colliders up to
today. The reaction was investigated by the \cite{VENUS} collaboration 1989 from energies $\sqrt{s}$ = 55 GeV - 57 GeV, \cite{OPAL}
collaboration 1991 at the $Z^0$ pole at $\sqrt{s}$ = 91 GeV, \cite{TOPAS} collaboration 1992 at $\sqrt{s}$ = 57.6 GeV, \cite{ALEPH}
collaboration 1992 at the Z$^0$ pole $\sqrt{s}$ = 91.0 GeV, \cite{DELPHI} collaboration from 1994 to 2000 at energies $\sqrt{s}$ = 91.0 GeV
to 202 GeV, \cite{L3} collaboration from 1991 to 1993 at the Z$^0$ pole range from $\sqrt{s}$ = 88.5 GeV - 93.7 GeV, \cite{L31} collaboration
2002 from $\sqrt{s}$ = 183 GeV - 207 GeV and \cite{OPAL1} collaboration 2003 from $\sqrt{s}$ = 183 GeV - 207 GeV. In summary cross sections of
the $e^+\,e^-\to \gamma\,\gamma$  reaction have been measured at centre of mass energies in total from $\sqrt{s}$ = 55 GeV to 207 GeV by these
six collaborations. Possible deviations from QED were studied in terms of contact $e\,e\,\gamma\,\gamma$ interaction and excited electron exchange.

The more recent experimental data obtained in \cite{HERA}, \cite{Tevatron}, \cite{ATLAS} and \cite{CMS} do not see evidence for the excited
leptons. Nevertheless, the future colliders with higher centre-of mass enrgy and luminosity will continue the search for the excited leptons.
The most recent experimental results on the excited electrons mass are obtained by the OPAL and the ATLAS collaborations \cite{OPAL1}.
The mass exclusion limits of the excited electrons are $m_{e^*}>$ 103.2 GeV for pair production ($e^+e^-\to e^*e^*$) and $m_{e^*}>$ 3000 GeV
for single production.

The large-angle two photon production in $e^+e^-$ annihilation was proposed as a possible process to monitor the luminosity of a future
$e^+e^-$ circular and linear colliders. Future colliders with polarized beams
such as (FCC-ee) \cite{CC19,AA19}, CEPC \cite{SG18}, ILC \cite{BB13} and CLIC \cite{ABD12} could provide the accuracy test of SM and probe the new physics signal.

In this paper we investigate the influence of the exited electron and contact $e\,e\,\gamma\,\gamma $ interaction on the angular distribution in the reaction
of the two--photon annihilation of $e^+e^-$-pair:
\begin{equation}\label{eq:reac1}
e^+(p_2)+e^-(p_1)\rightarrow \gamma (k_1)+\gamma (k_2).
\end{equation}
It was taken into account not only the contribution of the interference of these exotic mechanisms with standard QED amplitude
but also theirs contribution itself. The influence of the contact interaction on the polarization observables
in this reaction has been investigated for the cases of only one or both polarized beams. The excited electron can contribute
to the reaction (\ref{eq:reac1}) by its exchange in $t-$and $u-$channels. The effect of the excited electron contribution can be seen as distortion
in the angular distribution beyond the region of the forward or backward scattering. We investigate also the influence of the
excited electron on the polarization observables for the case of both polarized beams.

\section{Excited electron contribution}
\hspace{0.7cm}

The standard QED mechanism of the reaction (\ref{eq:reac1}) is described by two
Feynman diagrams (Fig. 1a,b). The production of the excited electron
in the intermediate state (in $t-$ and $u-$ channels) in this
reaction is described by two additional Feynman diagrams (Fig.
1\,a,b).

\begin{figure}
\centering
\includegraphics[width=0.45\textwidth]{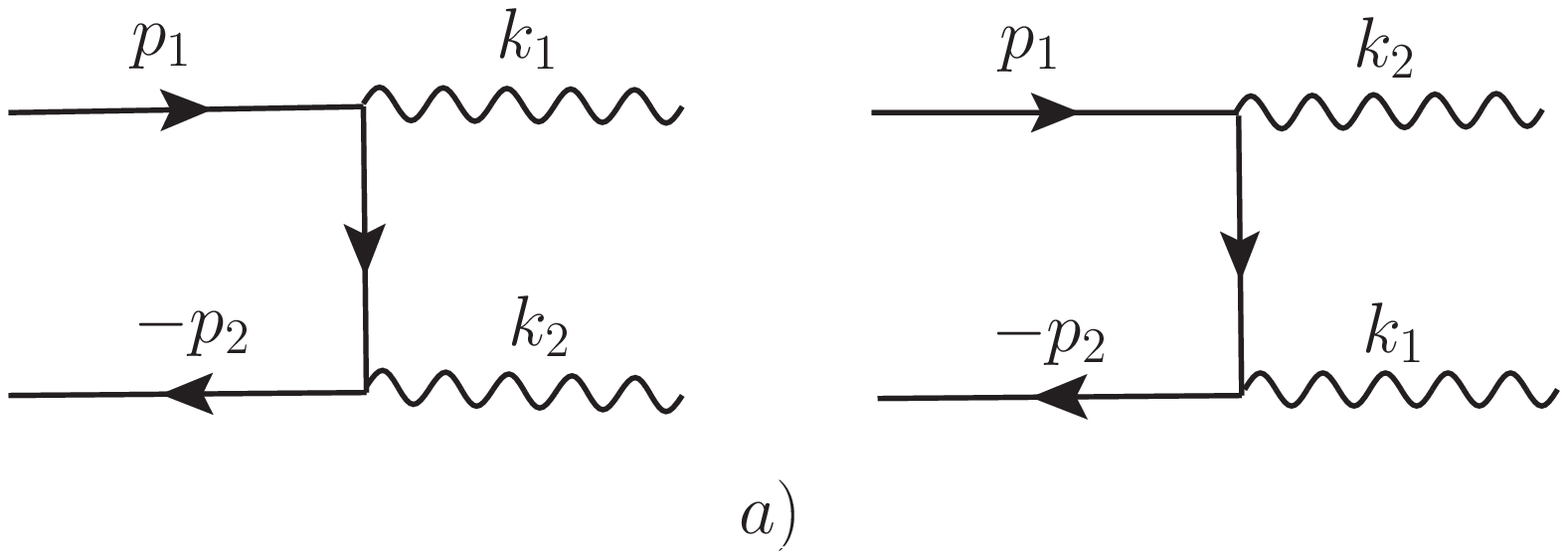}
\includegraphics[width=0.45\textwidth]{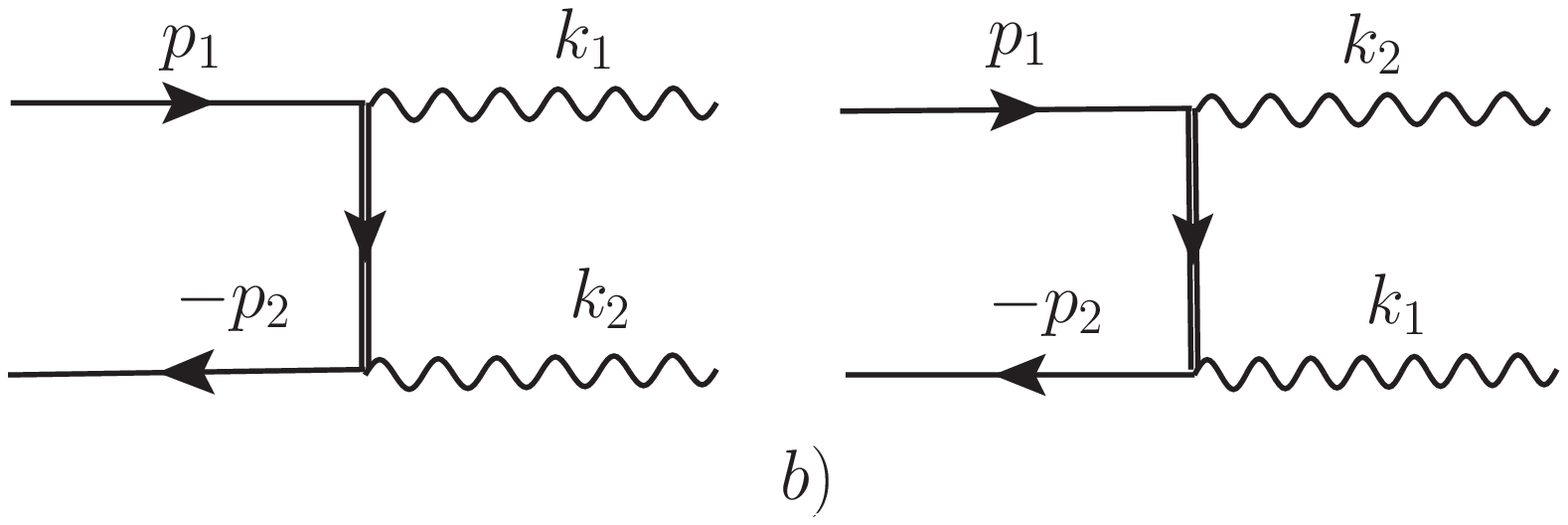}
\includegraphics[width=0.3\textwidth]{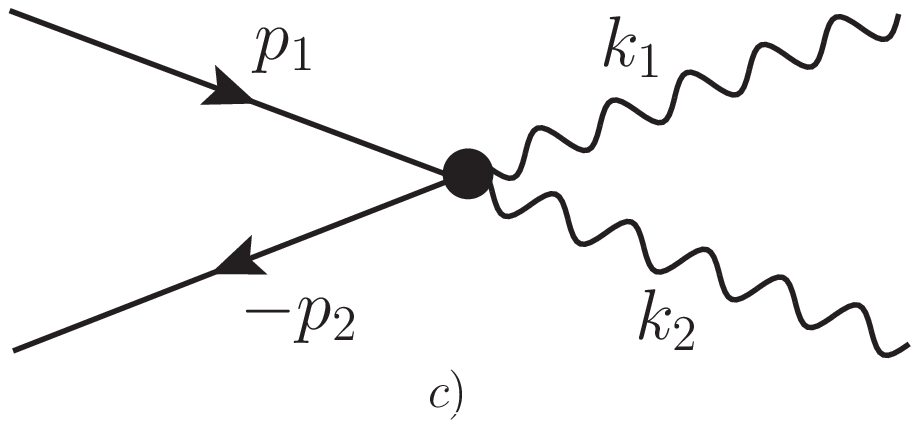}
 \parbox[t]{0.9\textwidth}{\caption{Feynman diagrams for the process $e^+\,e^-\rightarrow \gamma\,\gamma$. a)\,QED diagrams,
  b)\,diagrams with the exited electron in $t-$ and $u-$ channels, c)\, so-called contact diagram. }\label{fig.1}}
\end{figure}

The differential cross section of the reaction (\ref{eq:reac1}) after average over the spin states of the initial particles can be written as
follows
\begin{equation}\label{eq:ds2}
\frac{d\sigma}{d\Omega}= \frac{\omega}{W}\frac{|M|^2}{128\pi ^2}
[s(s-4m^2)]^{-\frac{1}{2}},
\end{equation}
where $\omega $ is the photon energy in CMS of the reaction (\ref{eq:reac1}), $W$ is the
total energy of the initial beams, $m$ is the electron mass, $s=W^2.$

The matrix element of the reaction (\ref{eq:reac1}) is the sum of two
contributions, namely: $M_{\gamma}$ (this part corresponds to the
pure QED mechanism and it is described by the Feynman diagrams
represented in Fig.\,1\,a) and $M_{ex}$ which corresponds to the
excited electron contribution (diagrams of Fig.\,1\,b).

Then the differential cross section of the reaction (\ref{eq:reac1}) can be
written, in this aproach, as
\begin{equation}\label{eq:dssum3}
\frac{d\sigma}{d\Omega}=\frac{d\sigma_{\gamma}}{d\Omega}+
\frac{d\sigma_{int}}{d\Omega}+\frac{d\sigma_{ex}}{d\Omega},
\end{equation}
where the first term is the cross section of this reaction
corresponding to the QED mechanism, the second one - contribution of
the interference of two mechanisms (QED and the excited electron)
and the last term - the contribution of the excited electron itself.

The matrix element $M_{\gamma}$ is well known
\begin{equation}\label{eq:Mg4}
M_{\gamma}=e^2\bar u(-p_2)[(t-m^2)^{-1}\hat A_2(\hat p_1-\hat k_1+m)
\hat A_1+(u-m^2)^{-1}\hat A_1(\hat p_1-\hat k_2+m)\hat A_2]u(p_1),
\end{equation}
where $e$ is the electron charge, $A_{1\mu}\,(A_{2\mu})$ is the polarization
4--vector  of the first (second) photon; $p_1\,(p_2),\, k_1\,(k_2)$ is the
4--momentum of the electron (positron), first (second) photon, respectively;
$t=(p_1-k_1)^2, \,u=(p_1-k_2)^2.$

At large energies (where at last and future time the experiments, investigating this
reaction, were done and will do) it is possible to neglect by the electron mass $m$ and
then the differential cross section of the reaction (\ref{eq:reac1}1), caused by the
standard QED mechanism (\ref{eq:dssum3}), has the form
\begin{equation}\label{eq:dsg5}
\frac{d\sigma_{\gamma}}{d\Omega}= \frac{\alpha^2}{s}
\frac{1+x^2}{1-x^2},
\end{equation}
where $x=cos\theta ,\,\, \theta$ is the angle between the electron and photon
3$-$momenta. The coordinate frame in CMS of the reaction (\ref{eq:reac1}) is chosen as: $z$
axis is directed along the initial electron 3$-$momentum and photon 3$-$momentum
lies in the $xz$ plane (the reaction plane). The expression (\ref{eq:dsg5}) is valid
for all angles except forward and backward scattering (i.e., $\theta=0^0,\,
180^0$), where it is necessary to take into account the electron mass in
the denominator.

We assume that the spin of the excited electron is 1/2 and its interaction
with the electromagnetic field is described by following effective
Lagrangian \cite{L65}
\begin{equation}\label{eq:Lag6}
L(ee^*\gamma)= \frac{e\lambda}{2M}\bar u_e\sigma_{\mu\nu}\bar u_{e^*}
F_{\mu\nu}+h.c.,
\end{equation}
where
$\sigma_{\mu\nu}=(\gamma_{\mu}\,\gamma_{\nu}-\gamma_{\nu}\,\gamma_{\mu})/2$,
$M$ is the excited electron mass, $\lambda $ is the dimensionless
coupling constant and $F_{\mu\nu}$ is the photon field-strength
tensor. Then the part of the matrix element, describing the
contribution of the excited electron to the reaction (\ref{eq:reac1}), can be
written as
\begin{equation}\label{eq:Mex7}
M_{ex}=-(\frac{e\lambda}{M})^2\bar u(-p_2)[(t-M^2)^{-1}\hat A_2\hat
k_2 (\hat p_1+M)\hat k_1\hat A_1+ (u-M^2)^{-1}\hat A_1\hat k_1(\hat
p_1+M)\hat k_2\hat A_2]u(p_1).
\end{equation}
As it was already mentioned above, at high energies one can neglect
the electron mass (we assume also that $M>>m$ and experimental data
suggest this assumption). Using expression (\ref{eq:Mex7}) as matrix element
$M_{ex}$ and expression (\ref{eq:Mg4}) for the matrix element $M_{\gamma}$ one
can obtain the following formula for the interference contribution
(in the used approximation)
\begin{equation}\label{eq:dsint8}
\frac{d\sigma_{int}}{d\Omega}= \frac{\alpha^2\lambda^2}{M^2}
[1-x^2+y(1+x^2)][(1+y)^2-x^2]^{-1},
\end{equation}
where $y=2M^2/s.$ In the same approximation the term, caused by the contribution
of the excited electron itself, has the form
\begin{equation}\label{eq:dsex9}
\frac{d\sigma_{ex}}{d\Omega}= \frac{\alpha^2\lambda^4}{8}
\frac{ys^2}{M^6}[4y^3+y^2(1-x^2)(9+x^2)+6y(1-x^2)^2+(1-x^2)^3]
[(1+y)^2-x^2]^{-2}.
\end{equation}

Let us do numerical estimations for the contribution of the excited
electron to the observables of the reaction (\ref{eq:reac1}). In the paper
\cite{A95}, on the basis of the analysis of the experimental data on
the differential cross section of the reaction (\ref{eq:reac1}), it was obtained
the following constraint on the excited electron mass: $M\ge 146$
GeV (assuming that the coupling constant $\lambda =1$).
Below in numerical estimations we use $M=150, \, 300$ GeV and $\lambda =1$ or $ 0.1$ at $W=200,\, 500$ GeV.

To estimate the influence of the exited electron on the angular distribution at different its parameters (mass and coupling constant) we plot
in Fig.\,2 the quantity $R_{ex}$ defined by Eq.\,(\ref{eq:Rex10})
\begin{equation}\label{eq:Rex10}
R_{ex}=\bigg(\frac{d\,\sigma_{int}}{d\,\Omega}+\frac{d\,\sigma_{ex}}{d\,\Omega} \bigg)/\frac{d\,\sigma_{\gamma}}{d\,\Omega}
\end{equation}

\begin{figure}
\centering
\includegraphics[width=0.45\textwidth]{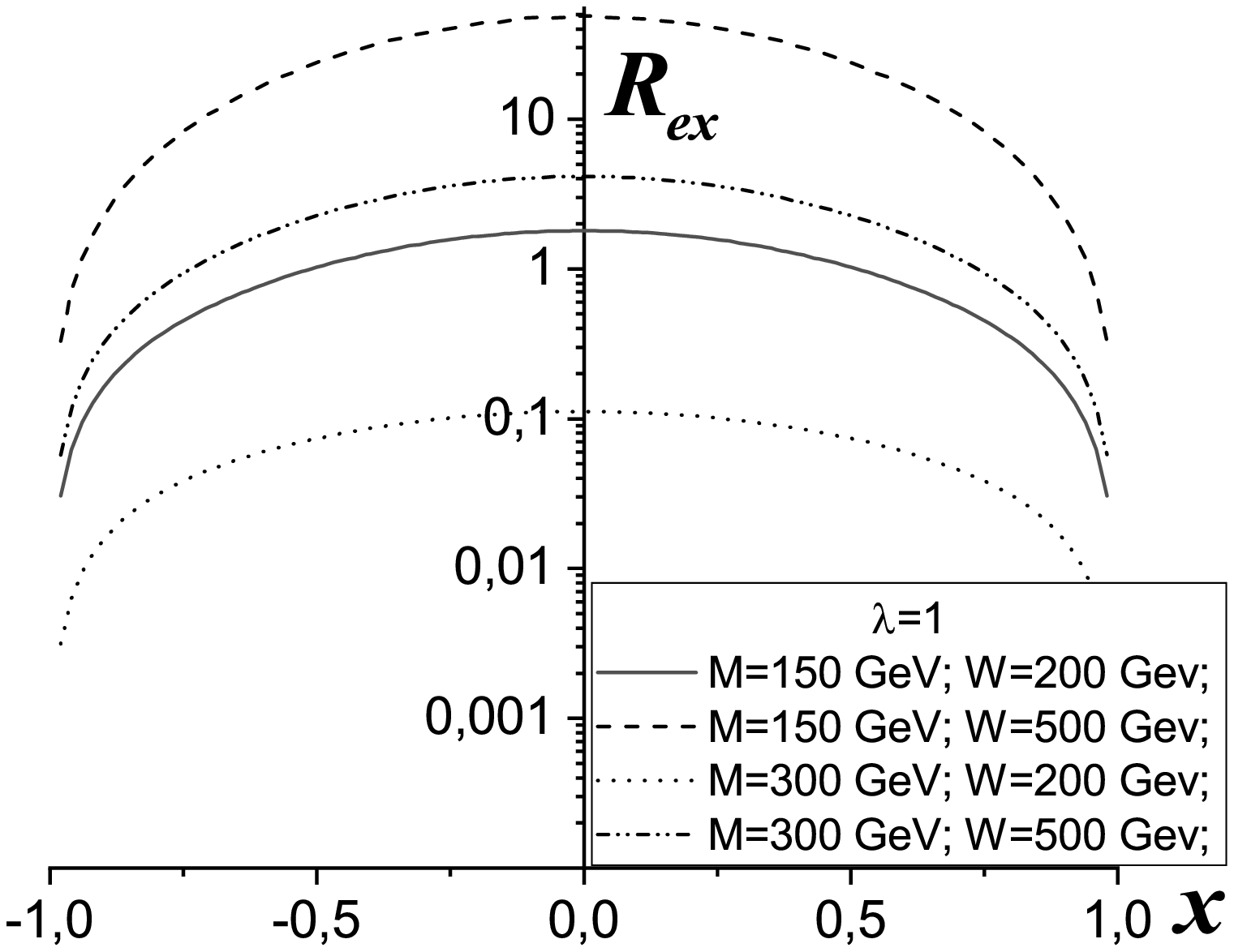}
\includegraphics[width=0.45\textwidth]{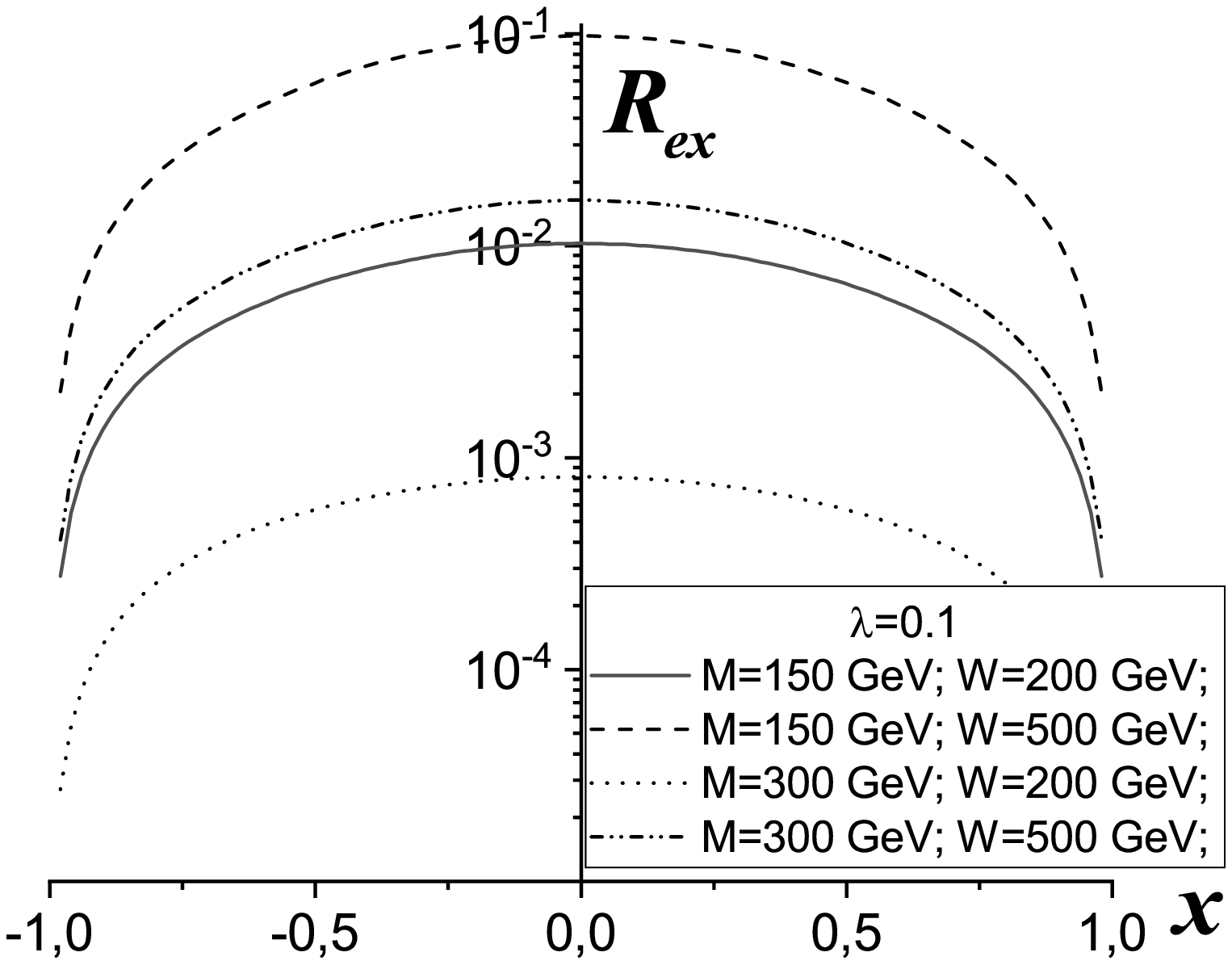}
 \parbox[t]{0.9\textwidth}{\caption{The angular dependence of the quantity  $R_{ex}$ defined by Eq.\,(\ref{eq:Rex10}) at different
 values of the exited electron parameters ($M,\,\lambda$) and the total c.m.s. energy $W$). }\label{fig.2}}
\end{figure}

 It is seen that at fixed beam energy the most
sensitivity to the contribution of the excited electron takes place
at the angles far from the region of the forward and backward scattering since
cross section, caused by the pure QED mechanism, has sharp peak in these cases.
 Note that in these angular region, where the sensitivity to the contribution of
the excited electron is maximal, the cross section decreases
appreciably and it requires more time to collect comparable
statistics. It is seen from the figures that at fixed value of the mass $M$ relative contribution
of the excited electron increases quickly with the increase of beam
energy (admittedly the value of the cross section decreases although
not so quickly).

It turns out that at
fixed beam energy the excited electron contribution to
the ratio $R_{ex}$ decreases strongly with the increase of the
mass $M.$  The sensitivity of the excited electron contribution to the coupling constant $\lambda$  (at fixed
values of $M$ and $W$) is very strong. For example, if we
reduce the $\lambda$   to one-tenth (from 1 to 0.1) the ratio
$R_{ex}$ reduces by a factor of 100.

Thus, the investigation of the
reaction (\ref{eq:reac1}), at future high energy linear electron-
positron colliders (CLIC or ILC with $W \simeq 500$ GeV),
can give more strict constraints on the excited electron
parameters (mass and coupling constant).

Let us consider the influence of the excited electron on the
polarization observables of the reaction (\ref{eq:reac1}) for the case when both
beams have arbitrary polarization. The mechanism, caused by the
exchange of the excited electron, does not lead to non--zero
polarization effects in the case when only one beam is polarized (at
least, in the lowest order of the perturbation theory) since the
reaction of the excited electron production conserve the space
parity (it is seen from the expression for  the Lagrangian (\ref{eq:Lag6}).

In the case when the initial beams have arbitrary polarization, the
differential cross section of the reaction (\ref{eq:reac1}) (taking into account the
contribution of the excited electron in the intermediate state) can be
written as
\begin{equation}\label{eq:gpol11}
\frac{d\sigma}{d\Omega}=\frac{d\sigma_0}{d\Omega}
(1+C_{zz}\xi_{1z}\xi_{2z}+C_{xx}\xi_{1x}\xi_{2x}+C_{yy}\xi_{1y}\xi_{2y}+
C_{zx}\xi_{1z}\xi_{2x}+C_{xz}\xi_{1x}\xi_{2z}),
\end{equation}
where $C_{ij}$ are the spin correlation coefficients, ${\vec \xi}_{1}\,
({\vec \xi}_{2})$ is the unit vctor along the electron (positron) polarization
in its rest system. Thus, $\xi_{iz}$ describes the longitudinal polarization
of the beams, and $\xi_{ix}\,(\xi_{iy})$  - the transverse polarization of the
beams and polarization vector lies in the reaction plane (orthogonal to the
plane). Let us note that $C_{xz},\, C_{zx}$ coefficients are proportional to
the electron mass and, therefore, they are zero in the high energy limit. The
rest coefficients in this limit have the form
$$\sigma_{0}C_{xx}=\frac{\alpha^2}{s}
\bigl\{1+\frac{\lambda^2s}{M^2}(1+y)(1-x^2)[(1+y)^2-x^2]^{-1}+
\frac{\lambda^4}{y^2}(1-x^2)^2[(1+y)^2-x^2]^{-2}\bigr\}, $$
\begin{equation}\label{eq:polK12}
C_{yy}=-C_{xx},
\end{equation}
$$\sigma_{0}C_{zz}=\frac{\alpha^2}{s}
\bigl\{(1+x^2)(1-x^2)^{-1}+\frac{2\lambda^2}{y}[1-x^2+y(1+x^2)]
[(1+y)^2-x^2]^{-1}+ $$
$$+\frac{\lambda^4}{y^2}[-4y^3+y^2(1-x^2)(x^2-7)-2y(1-x^2)^2+(1-x^2)^3]
[(1+y)^2-x^2]^{-2}\bigr\}, $$
where $\sigma_0=d\sigma_0/d\Omega$ is the differential cross section of the
reaction (\ref{eq:reac1}) for the case when all particles are unpolarized. Some formulae
for the differential cross section and polarization observables of the
reaction (\ref{eq:reac1}) (taking into account the electron mass) are given in the
Appendix A.

\begin{figure}
\centering
\includegraphics[width=0.45\textwidth]{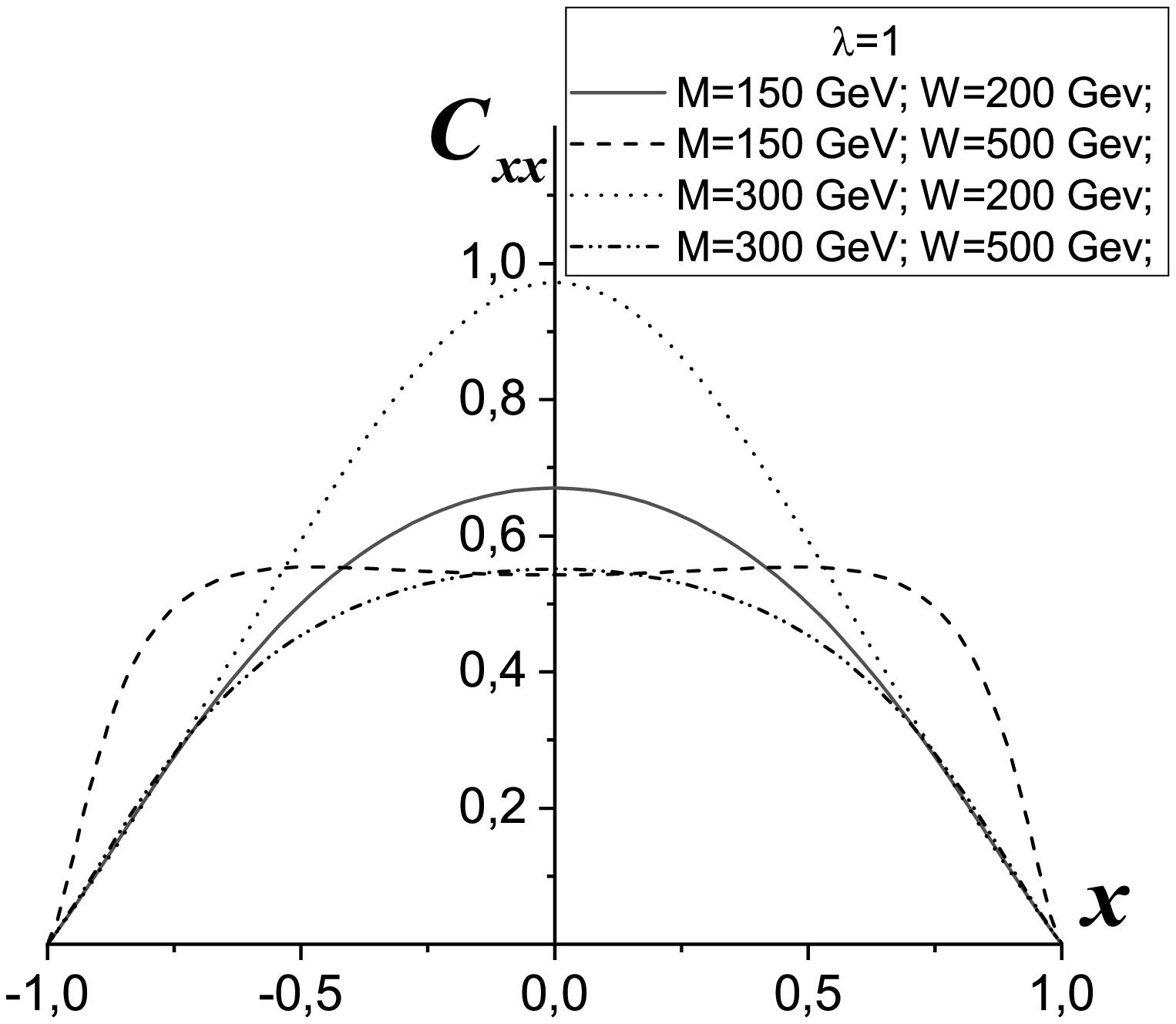}
\includegraphics[width=0.45\textwidth]{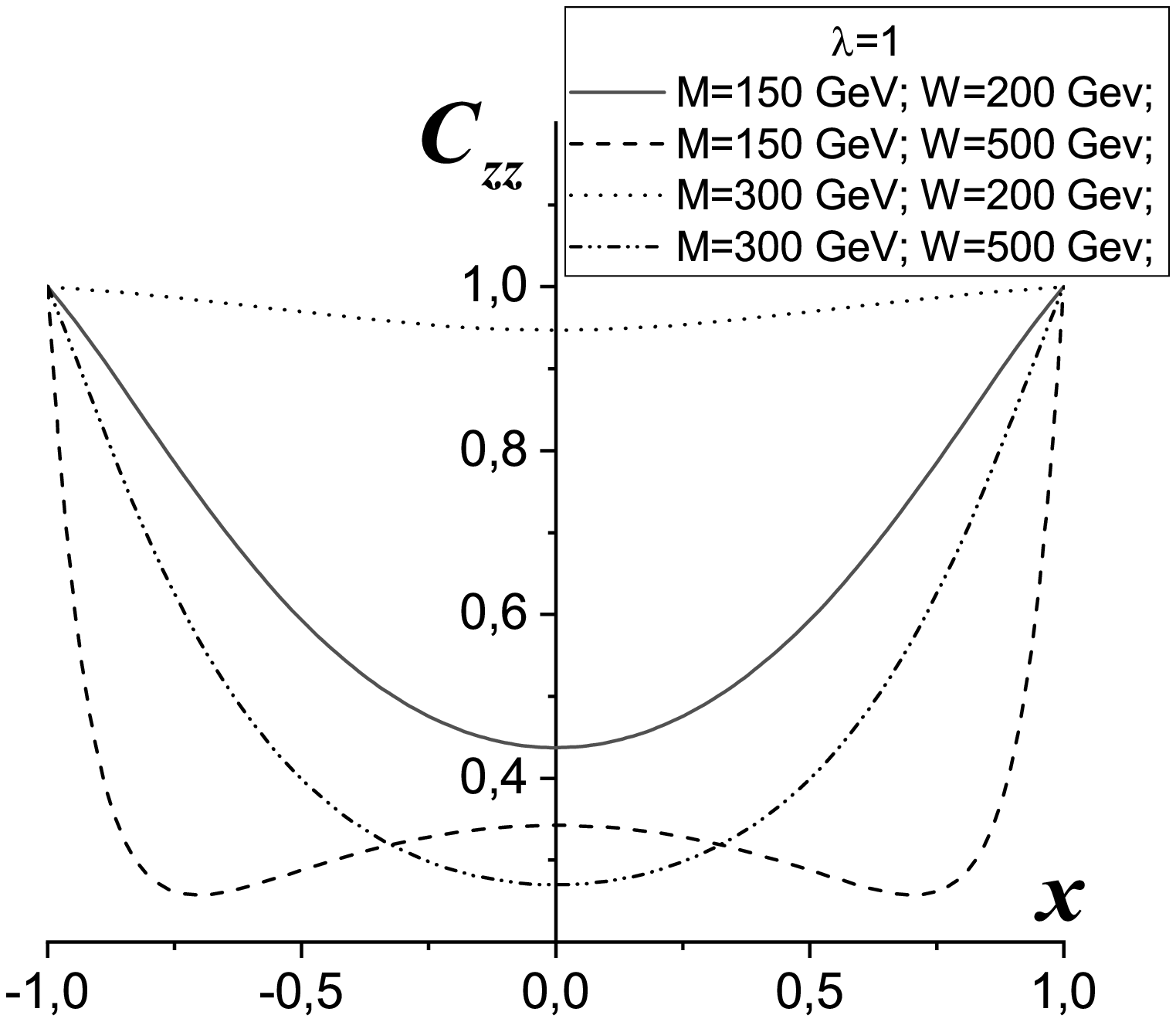}
 \parbox[t]{0.9\textwidth}{\caption{The angular dependence of the correlation coefficients $C_{xx}$ and $C_{zz}$
 for polarized beams at different exited electron mass and the total c.m.s. energy.}\label{fig.3}}
\end{figure}

We did the numerical estimations of the spin correlation coefficients $C_{xx}$ and $C_{zz}.$  They are given in
Fig.\,3. The analysis of their behaviour depending on the excited electron mass, the total energy of the beams was
done in the same way as for the ratio $R_{ex}$. One can see from Fig.\,3 that the spin correlation coefficients have
appreciable values over a wide range of the scattering
angles (especially for the $C_{zz}$ quantity).

Note that the coefficient $C_{zz}=1$ for the pure QED mechanism. Such behaviour of the spin correlation
coefficients is important since at small angles the cross section is large and this circumstance permits to collect more
data. The spin correlation coefficients do not change strongly with increasing of the total energy $W.$ The de
dependence of these coefficients on the excited electron mass reduces with the increase of the energy $W.$
Note that the proposed future linear colliders ILC and CLIC cover, at first run, the region $W\simeq 500$ GeV
(with longitudinal beam polarization). Both colliders have possible upgrades to 1 (ILC) and 3 TeV (CLIC) \cite{BJ19}.


\section{Contact $ee\gamma\gamma $ interaction}
\hspace{0.7cm}

Earlier, the contact $e\,e\gamma\,\gamma $ interaction (diagram c) in Fig./,1) was investigated
in the papers \cite{ENN, DT85} where the contribution of this
contact interaction was taken into account to the reaction (\ref{eq:reac1}).
Besides, in the paper \cite{DT85} it was considered the
manifestation of this contact interaction in the reactions
$e^+e^-\to 3\gamma ,$ $\gamma +F\to F+l\bar l$ (where $F$ designates
arbitrary fermion). In these papers the contribution of contact
interaction was taken into account on the level of its interference
with standard QED mechanism. The manifestation of a possible contact
$q\bar\, q\gamma\,\gamma $ interaction (independent of the flavour of
the quark $q$) in the reactions at the hadron colliders was
investigated in the paper \cite{R94}. In this paper, the
contribution of the contact interaction to the cross section of the
reaction (\ref{eq:reac1}) is completely taken into account. The influence of the
initial particle polarizations on the observable characteristics of
these processes was not investigated in these papers.

The experimental search for various types of the contact interaction
are being done at present at the lepton, lepton--hadron and pure
hadron colliders. Thus, the contact interaction were searched by
means of measuring the differential cross sections of the
electromagnetic processes $e^+\,e^-\to e^+\,e^-, \mu^+\,\mu^- ,
\tau^+\,\tau^- $ \cite{B93}, $e^+\,e^-\to \gamma\,\gamma $ \cite{A95, B93}
$e^+\,e^-\to 3\gamma $ \cite{A95}, and also deep inelastic $e^{\pm}p $
scattering \cite{A94} (the references on earlier sources about the
experimental search for these contact interactions see in these
papers). Using the results of various experiments it was obtained
the lower limits for the corresponding energy scales $\Lambda .$

The matrix element of the reaction (\ref{eq:reac1}) is the sum of $M_{\gamma}$
(the pure QED mechanism) and $M_{ci}$ which describes the
contribution of the contact $ee\gamma\gamma $ interaction.
(it corresponds to the Feynman diagram shown in Fig.\,1c)).
Then the differential cross section of the reaction (\ref{eq:reac1}) can be written, in
this approach, as a sum of three contributions, namely, by Eq.\,(\ref{eq:dssum3}),
where instead of $ex$ index it is necessary to put $ci$ index (the
contact interaction contribution). Then, in this case, the second
term describes the contribution of the interference of two
mechanisms (QED and the contact interaction), and the last one - the
contribution to the cross section of the contact interaction itself.

The effective Lagrangian of any contact interaction is constructed
using the fields of particles known at present and is proportional
to the lowest possible power of $1/\Lambda $ which depends on the
dimensionality of the fields entering the Lagrangian. When
constructing this Lagrangian, we demand that fermion currents
corresponding this Lagrangian conserve the helicity. This assumption
is necessary, for example, for various types of the composite
models. This condition ensures that masses of known particles are
much less than the energy scale $\Lambda.$ Various possible choices
of the helicity of the fields participating in the construction of
the contact interaction Lagrangian lead to different predictions for
the angular distributions (and also for the polarization
observables) in the reactions where the contribution of the contact
interaction is taken into account.

The contact interaction for two fermions and two bosons was
considered, in general case, in the paper \cite{MPR}. For the case
of the $ee\gamma\gamma $, the Lagrangian of the contact interaction
can be written as
\begin{equation}\label{eq:LagC13}
L(ee\gamma\gamma )=2\frac{ie^2}{\Lambda^4}F_{\mu\sigma}F_{\nu\sigma}
(\eta_L\bar \psi_L\gamma_{\mu}d_{\nu}\psi_L+
\eta_R\bar \psi_R\gamma_{\mu}d_{\nu}\psi_R),
\end{equation}
where $d_{\nu}$ is the derivative, $\psi $ is the electron wave
function, $F_{\mu\nu}$ is the strength of the electromagnetic field,
indices $L$ and $R$ designate the helicity of the electron field,
namely: $\psi_{L,\,R}=(1\mp \gamma_5)\psi/2.$ The dimensions of the
fields participating in the effective Lagrangian (\ref{eq:LagC13}) lead to the
fact that it is proportional to $\Lambda^{-4}.$ The dimensionless
coefficients $\eta_i\, (i=L,\,R)$ show the helicity of the electron
current. The particular model of the contact $e\,e\,\gamma\,\gamma
$ interaction is determined by the set of the dimensionless
parameters $\eta_i$ and corresponding energy scale $\Lambda$. The
transition amplitude corresponding to the contact interaction is
real value and can be both positive and negative (and each sign of
the parameter $\eta_i$ is associated with different value of the
energy scale $\Lambda$). Thus, the effective Lagrangian (\ref{eq:LagC13})
determine $4\times 2$ various models of the contact interaction
depending on the choice of the electron current helicity (see the
Table\,\,I).

\begin{table*}
\begin{tabular}{|c|c|c|}
\hline
Model & $\eta_R$  & $\eta_L$\\
\hline
L & 0 & $\pm$ 1\\
\hline
R & $\pm$ 1 & 0\\
\hline
L+R & $\pm$ 1 & $\pm$ 1\\
\hline
L-R & $\pm$ 1 & $\mp$ 1\\
\hline
\end{tabular}
\caption{The models determined by the Lagrangian (\ref{eq:LagC13}).}
\end{table*}

The matrix element $M_{ci}$, corresponding to the effective
Lagrangian (\ref{eq:LagC13}), can be written in the following general form (in
the impulse representation)
\begin{equation}\label{eq:Mci14}
M_{ci}=-\frac{e^2}{\Lambda^4}V_{\mu}\bar u(-p_2)O_{\mu}u(p_1),
\end{equation}
where
$V_{\mu}=p_{1\nu}\,(F_{1\mu\sigma}\,F_{2\nu\sigma}+1\Leftrightarrow 2)$,
$O_{\mu}=\gamma_{\mu}\,(a+b\gamma_5),$ $F_{1\mu\nu}\,(F_{2\mu\nu})$ is
the tensor of the electromagnetic field of the first (second)
photon. Various versions of the models from the Table\,I are
corresponded to the following sets of the constants $a$ and $b$:
model $L$ corresponds to $a=b=\pm 1$; model $R$ - $a=-b=\pm 1$;
model $(L+R)$ - $a=\pm 2, b=0$; model $(L-R)$ - $a=0, b=\pm 2$.

For the interference contribution (between QED and the contact
interaction mechanisms) to the differential cross section of the
reaction (\ref{eq:reac1}) one can obtain the following simple expression valid at
high energies
\begin{equation}\label{eq:DsCint15}
\frac{d\sigma_{int}}{d\Omega}=a\frac{\alpha^2s}{4\Lambda^4}
(1+x^2).
\end{equation}

From this formula one can see that in the case of $(L-R)$ model the
interference contribution is zero (in this case the constant $a=0$),
since $O_{\mu}\approx \gamma_{\mu}\,\gamma_5,$ and the amplitude
corresponding to the QED mechanism conserves parity. The
interference contribution for the $(L+R)$ model twice as large than
similar value for $L$ or $R$ model. In the same approximation
($m=0$), the term corresponding to the contact interaction mechanism
itself has the form
\begin{equation}\label{eq:DsC16}
\frac{d\sigma_{ci}}{d\Omega}=\frac{\alpha^2s^3}{64\Lambda^8}
(a^2+b^2)(1-x^4).
\end{equation}
This formula show that the contact interaction contribution  for the case
of $(L\pm R)$ models twice as large than for the $L$ or $R$ models.

Let us consider the influence of the contact interaction on the
polarization observables in the reaction (\ref{eq:reac1}) for the case when
initial particles are polarized. Since in the general case the
contact interaction violates the space parity (constant $b\neq 0$)
then non--zero observables arise in the case when only one beam is
polarized. The pure QED mechanism of this reaction (at least without
taking into account the radiative corrections) does not lead to such
polarization effects. The electroweak corrections (at one--loop
level see, for example, \cite{BDK22}) can lead to additional term in the amplitude of this process
which violate the parity and, therefore, can lead to the non--zero
polarization observables.

Let only the electron beam is polarized. Then, in the limit of high energy,
the part of the interference contribution which is caused by the electron
beam polarization has the form (remember that QED part of the cross section
does not contain the polarization contribution)
\begin{equation}\label{eq:DsPint17}
\frac{d\sigma_{int}(\xi )}{d\Omega}=-b\frac{\alpha^2s}{4\Lambda^4}
(1+x^2)\xi_z,
\end{equation}
where ${\vec \xi}$ is the unit vector along the polarization of the
electron in its rest system. From this formula one can see that the
effect is not zero only when the beam has longitudinal polarization,
i.e., when the beam polarization vector is directed along (opposite)
beam momentum. Note that this result valid in the high energy limit.
The transverse polarization of the beam leads to the terms which are
suppressed by small factor ($m/W$). Naturally that for the case of
the $(L+R)$ model the contribution (\ref{eq:DsPint17}) turns into zero (since
constant $b=0$). The reason is that the contact interaction in this
case conserves the parity.

The part of the cross section of the reaction (\ref{eq:reac1}), caused by the contribution
of the contact interaction itself, depends on the electron beam polarization
(in the high energy limit) by the following way
\begin{equation}\label{eq:DsPci18}
\frac{d\sigma_{ci}(\xi
)}{d\Omega}=-ab\frac{\alpha^2s^3}{32\Lambda^8} (1-x^4)\xi_z.
\end{equation}
From this formula one can see that this term turns into zero for the
$(L\pm R)$ models. The remarks to Eq.\,(\ref{eq:DsPint17}) are valid also for this formula.

When one of the beams has longitudinal polarization one can introduce additional
polarization observable which is the most sensitive to new contact interactions.
It is the so--called left-right asymmetry determined as
\begin{equation}\label{eq:DsPlr19}
A_{LR}=\frac{\sigma_L-\sigma_R}{\sigma_L+\sigma_R},
\end{equation}
where $\sigma_{L,R}$ is the differential cross section of the reaction (\ref{eq:reac1})
in the case when the electron beam is left-- ($\xi_z=-1$) or right--
($\xi_z=+1$)polarized and the positron beam is not polarized. One can show
that in the high energy limit this asymmetry is
\begin{equation}\label{eq:DsPlr20}
A_{LR}=\frac{2bR(1-x^2)[8+\epsilon aR(1-x^2)]}
{64+16aR(1-x^2)+\epsilon R^2(a^2+b^2)(1-x^2)^2},
\end{equation}
where $R=s^2/\Lambda^4$. At $\epsilon =0$ this expression describes
the asymmetry in the approximation when the contribution of the main
QED mechanism and its interference with the contact interaction
mechanism is taken into account. At $\epsilon =1$ this expression
describes the asymmetry in the approximation when contribution of
the contact interaction mechanism is completely taken into account.

Let us consider the case when both beams have arbitrary polarization. Note
that synchrotron radiation in the electron--positron colliders leads to the
polarization of these beams. This polarization is transverse and its value
is appreciable.

The pure QED mechanism lead to the following contribution to the differential
cross section of the reaction (\ref{eq:reac1}) which depends on the polarization of both
beams (in the high energy limit)
\begin{equation}\label{eq:DsPQED21}
\frac{d\sigma_{\gamma}(\xi_1, \xi_2 )}{d\Omega}=
\frac{d\sigma_{\gamma}}{d\Omega}[1+\xi_{1z}\xi_{2z}+
\frac{1-x^2}{1+x^2}(\xi_{1x}\xi_{2x}-\xi_{1y}\xi_{2y})].
\end{equation}

The interference (between QED and the contact interaction
mechanisms) contribution to the differential cross section of the
reaction (\ref{eq:reac1}) caused by the polarization of both beams has the
following form (in the high energy limit)
\begin{equation}\label{eq:DsP2int22}
\frac{d\sigma_{int}(\xi_1, \xi_2 )}{d\Omega}=
a\frac{\alpha^2s}{4\Lambda^4}
[(1-x^2)(\xi_{1x}\xi_{2x}-\xi_{1y}\xi_{2y})+
(1+x^2)\xi_{1z}\xi_{2z}].
\end{equation}

This expression is valid for the case when constants $a$ and $b$ (describing
the effective Lagrangian of the contact interaction) are real quantities. If
the constant $b$ become complex value (for example, due to the contribution
of the radiative corrections or presence of the imaginary part in the
propagator of the supermassive gauge bosons that lead to the contact
interaction in the low energy limit) the expression (\ref{eq:DsP2int22}) acquires additional
contribution:
$$\frac{\alpha^2s}{4\Lambda^4}(x^2-1)(\xi_{1x}\xi_{2y}+
\xi_{1y}\xi_{2x})Im(b). $$

The manifestations of various effects caused by the imaginary part of the
effective coupling constant, in the reactions on colliding electron--positron
beams, were investigated in detail in Ref. \cite{MT95}. In this paper it was
shown that for the existing level of the experimental accuracy one can not
neglect, for some asymmetries, by the imaginary parts of the effective
coupling constants. Note also that the interference contribution (\ref{eq:DsP2int22}) turns
into zero for the case of the $(L-R)$ model.

The term in the differential cross section of the reaction (\ref{eq:reac1}),
caused by the contribution of the contact interaction itself and depending
on the polarization of both beams, has the following form
(in the high energy limit)
\begin{equation}\label{eq:DsP2ci23}
\frac{d\sigma_{CI}(\xi_1, \xi_2 )}{d\Omega}=
\frac{\alpha^2s^3}{64\Lambda^8}
[(a^2-b^2)(1-x^2)^2(\xi_{1x}\xi_{2x}-\xi_{1y}\xi_{2y})+
(a^2+b^2)(1+x^4)\xi_{1z}\xi_{2z}].
\end{equation}
From this formula one can see that in the high energy limit the transverse
polarization of the initial beams does not contribute to the cross section
for the case of the $L$ or $R$ models.

\section{Conclusion}
\hspace{0.5cm}

We have analyzed the influence of some mechanisms, which are beyond
the Standard Model framework, on the observables of the reaction
$e^+\,e^-\to \gamma\,\gamma $. The influence of the contact
$e\,e\,\gamma\,\gamma $ interaction and the presence of the excited
electron on the angular dependence of the differential cross section
and spin correlation coefficients (when both initial beams are
polarized) have been investigated for the reaction (\ref{eq:reac1}). These
effects turn out to be appreciable and their magnitude increases
quickly when initial beam energy grows. Therefore, the experimental
investigation of this reaction on the future
lepton colliders may essentially progress to limit the parameters of
the mechanisms under consideration.

\begin{center}
{\Large{\bf{Appendix A}}}
\end{center}
\setcounter{equation}{0}
\def\theequation{A.\arabic{equation}}

In this Appendix we present the exact expressions (taking into account the
electron mass) for the differential cross section of the $e^+e^-\to
\gamma\gamma $ reaction including the contribution of the excited electron
in the intermediate state.

The part of the cross section caused by the pure QED mechanism has the
following form in terms of the invariant variables (the averaging over the
spins of the initial particles was done)
\begin{equation}\label{1}
\frac{d\sigma_{\gamma}}{d\Omega}=\frac{\alpha^2}{2}
[s(s-4m^2)]^{-\frac{1}{2}}\bigl\{(u-m^2)(t-m^2)^{-1}+
(t-m^2)(u-m^2)^{-1}-4m^2(t-m^2)^{-1}-
\end{equation}
$$-4m^2(u-m^2)^{-1}-4m^4[(t-m^2)^{-1}+(u-m^2)^{-1}]^2\bigr\}, $$
where the invariant variables are defined as $t=(p_1-k_1)^2,$
$u=(p_1-k_2)^2,$ $s=(p_1+p_2)^2.$ In term of the angular variable this
expression is
\begin{equation}\label{2}
\frac{d\sigma_{\gamma}}{d\Omega}=\frac{\alpha^2}{\beta s}
\bigl [\frac{1+\beta^2(1+sin^2\theta )}{1-\beta^2cos^2\theta }-
\frac{2\beta^4sin^4\theta }{(1-\beta^2cos^2\theta )^2}\bigr ],
\end{equation}
where $\beta =p/E, $ $p(E)$ is the momentum (energy) of the initial
electron, $\theta $ is the angle between electron and one of the photons
momenta.

The interfernce part of the cross section has the following form in
terms of the invariant variables
\begin{equation}\label{3}
\frac{d\sigma_{int}}{d\Omega}=-\frac{\alpha^2\lambda^2}{M^2}
[s(s-4m^2)]^{-\frac{1}{2}}\bigl\{[2m^2s+(t-m^2)^2](t-M^2)^{-1}+
[2m^2s+(u-m^2)^2]
\end{equation}
$$(u-M^2)^{-1}+m^2s(s+m^2-mM)[(t-m^2)^{-1}(u-M^2)^{-1}+
(u-m^2)^{-1}(t-M^2)^{-1}]+ $$
$$+m^3(m-M)s[(t-m^2)^{-1}(t-M^2)^{-1}+(u-m^2)^{-1}(u-M^2)^{-1}]
\bigr\}, $$
where $M$ is the excited electron mass. In term of the angular variable
this expression is
\begin{equation}\label{4}
\frac{d\sigma_{int}}{d\Omega}=-\frac{\alpha^2\lambda^2}{\beta M^2}
\bigl \{(R_2-1)(1+2R_1)+(1+R_2)\beta^2cos^2\theta -R_1(1-
\beta^2cos^2\theta )^{-1}
\end{equation}
$$[(R_2-1)(2+R_1-RR_1)+2\beta^2cos^2\theta ]\bigr\}
[(R_2-1)^2-\beta^2cos^2\theta ]^{-1}, $$
where $R=M/m,$ $R_1=m^2/E^2,$ $R_2=(m^2-M^2)/2E^2.$

The part of the cross section caused by the excited electron
exchange in the $t-$ channel has the following form in terms of the
invariant variables
\begin{equation}\label{5}
\frac{d\sigma_{CI}}{d\Omega}=\frac{\alpha^2\lambda^4}{2M^4}
[s(s-4m^2)]^{-\frac{1}{2}}\bigl\{2(m^2+M^2)s(ut-m^4)
(t-M^2)^{-1}(u-M^2)^{-1}+
\end{equation}
$$+(ut-m^4+M^2s)[t-m^2)^{2}(t-M^2)^{-2}+
(u-m^2)^{2}(u-M^2)^{-2}]\bigr\}. $$
In term of the angular variable this expression is
\begin{equation}\label{6}
\frac{d\sigma_{CI}}{d\Omega}=\frac{\alpha^2\lambda^4}{\beta M^4}
\frac{s}{4}(R_2^2+R^2R_1+\beta^2sin^2\theta )^{-1}
\bigl \{R^2R_1^2+\beta^2sin^2\theta [4(R_1-R_2)+
\beta^2sin^2\theta ]+
\end{equation}
$$+2R_2^2\beta^2cos^2\theta (R^2R_1+\beta^2sin^2\theta )
(R_2^2+R^2R_1+\beta^2sin^2\theta )^{-1}\bigr\}. $$

\begin{center}
{\Large{\bf{Appendix B}}}
\end{center}
\setcounter{equation}{0}
\def\theequation{B.\arabic{equation}}

Here we present the helicity amplitudes of the $e^+\,e^-\to
\gamma\,\gamma $ reaction which are caused by following contributions: the
standard QED mechanism, the heavy electron excitation and contact
$ee\gamma\gamma $ interaction.

Let us define the helicity amplitudes as:
$$H=<\lambda_{\gamma_1},
\lambda_{\gamma_2}|M|\lambda_{e^+}, \lambda_{e^-}>, $$
where $\lambda_{\gamma_1}
\,(\lambda_{\gamma_2}), \lambda_{e^+}\,(\lambda_{e^-})$ are the helicity of the
first (second) photon, positron (electron), correspondingly. Altogether
there are 16 helicity amplitudes and the condition of the $P$--invariance
reduces this number to 8 independent amplitudes. But since we consider the
contact interaction of the general type with possible space parity violation
we write down all helicity amplitudes. Let us denote the helicity amplitudes
as
$$H_{1,\,16}=<\pm\,\pm|M|\pm\,\pm>, \ \ H_{5,\,12}=<\pm\,\mp|M|\pm\,\pm>, $$
\begin{equation}\label{B.1}
H_{2,\,15}=<\pm\,\pm|M|\pm\,\mp>, \ \ H_{6,\,11}=<\pm\,\mp|M|\pm\,\mp>,
\end{equation}
$$H_{3,\,14}=<\pm\,\pm|M|\mp\,\pm>, \ \ H_{7,\,10}=<\pm\,\mp|M|\mp\,\pm>, $$
$$H_{4,\,13}=<\pm\,\pm|M|\mp\mp>, \ \ H_{8,\,9}=<\pm\,\mp|M|\mp\,\mp>, $$
where for the final state ($\gamma\,\gamma $ system) the signs $\pm $ mean the
photon helicity $\pm 1$, and for the initial state ($e^+\,e^- $ system) the
signs $\pm $ mean the helicity $\pm 1/2$ of the electron or positron,
respectively.

The helicity amplitudes $H_i$ describing the standard QED mechanism have
following form (including the electron mass, i. e., $m\ne 0$)
$$H_i=-\frac{2e^2}{1-\beta^2cos^2\theta }h_i, $$
where
$$h_1=h_{16}=\gamma^{-1}(1-\beta ), h_2=h_{15}=h_3=h_{14}=0, $$
\begin{equation}\label{B.2}
h_4=h_{13}=-\gamma^{-1}(1+\beta ), h_5=h_{12}=h_8=h_{9}=
\gamma^{-1}\beta sin^2\theta ,
\end{equation}
$$h_6=-h_{11}=-\beta sin\theta (1+\cos\theta ),
h_{10}=-h_{7}=\beta sin\theta (1-\cos\theta ), $$
where $\theta $ is the angle between the electron and photon momenta,
$\beta =p/E, $ $\gamma =E/m, $ $p\,(E)$ is the electron momentum (energy).

At high energies (when the electron mass can be neglected) these amplitudes
are essentially simplified and we have
\begin{equation}\label{B.3}
H_6=-H_{11}=2e^{2}(1+cos\theta )/sin\theta, \ \ H_{10}=-H_{7}=-2e^{2}(1-cos\theta )/sin\theta.
\end{equation}
The rest amplitudes are zero in this approximation.

The helicity amplitudes $H_i$ describing the contribution of the heavy
electron to the $e^+\,e^-\to \gamma\,\gamma $ reaction (interaction $e^*\,e\,\gamma $
is described by the expression (\ref{eq:Lag6})) have following form (including the
electron mass)
$$H_i=e^2\lambda^2\frac{s}{M^2}h_i, $$
$$h_1=h_{16}=M(E+p)\Big [\frac{1-cos\theta }{t-M^2}+
\frac{1+cos\theta }{u-M^2}\Big ], $$
\begin{equation}\label{B.4}
h_4=h_{13}=-M(E-p)\Big [\frac{1+cos\theta }{t-M^2}+
\frac{1-cos\theta }{u-M^2}\Big ],
\end{equation}
$$h_2=-h_{15}=-m\,M\,sin\theta\Big [\frac{1}{t-M^2}-\frac{1}{u-M^2}\Big ], $$
$$h_3=-h_{14}=m\,M\,sin\theta\Big [\frac{1}{t-M^2}-\frac{1}{u-M^2}\Big ], $$
$$h_5=h_{12}=h_8=h_9=m\,p\,(sin\theta)^2\Big [\frac{1}{t-M^2}
+\frac{1}{u-M^2}\Big ], $$
$$h_6=-h_{11}=-p\,sin\theta (1+cos\theta )\Big [\frac{E-p}{t-M^2}
+\frac{E+p}{u-M^2}\Big ], $$
$$h_7=-h_{10}=-p\,sin\theta (1-cos\theta )\Big [\frac{E+p}{t-M^2}
+\frac{E-p}{u-M^2}\Big ], $$
where $M$ is the heavy electron mass.

At high energies these amplitudes are essentially simplified and we have
$$h_1=h_{16}=M\sqrt{s}\Big [\frac{1-cos\theta }{t-M^2}+
\frac{1+cos\theta }{u-M^2}\Big ], $$
\begin{equation}\label{B.5}
h_6=-h_{11}=-\frac{s}{2}\,sin\theta \frac{1+cos\theta }{u-M^2}, \ \
h_7=-h_{10}=-\frac{s}{2}\,sin\theta \frac{1-cos\theta }{t-M^2}.
\end{equation}
The rest amplitudes are zero in this approximation.

The helicity amplitudes $H_i$ describing the contribution of the contact
$e\,e\,\gamma\,\gamma $ interaction (see the expression (\ref{eq:LagC13})) have following form
(including the electron mass)
$$H_i=\frac{e^2}{2}\frac{s^{3/2}}{\Lambda^4}h_i, $$
$$h_1=h_{13}=m(\beta a+b),\ \  h_4=h_{16}=m(\beta a-b), $$
\begin{equation}\label{B.6}
h_5=h_9=h_8=h_{12}=-m\,\beta a\,(sin\theta)^2, \ \
h_2=h_3=h_{14}=h_{15}=0,
\end{equation}
$$h_6=p\,sin\theta (1+cos\theta )(a+\beta b), \ \
h_{10}=-p\,sin\theta (1-cos\theta )(a+\beta b),  $$
$$h_7=p\,sin\theta (1-cos\theta )(a-\beta b), \ \
h_{11}=-p\,sin\theta (1+cos\theta )(a-\beta b).  $$

At high energies we have
$$H_i=-\frac{e^2}{4}\frac{s^{2}}{\Lambda^4}\,sin\theta \,h_i, $$
where we introduce
$$h_6=(1+cos\theta )(a+b), \ \ h_{10}=-(1-cos\theta )(a+b),  $$
\begin{equation}\label{B.7}
h_7=(1-cos\theta )(a-b),\ \  h_{11}=-(1+cos\theta )(a-b).
\end{equation}
The rest amplitudes are zero in this approximation.

It can be shown also that for the $L$ model amplitudes $h_7=h_{11}=0$, and
for the $R$ model amplitudes $h_6=h_{10}=0$. For the $(L+R)$ model we have
$h_6=-h_{11}$, and $h_7=-h_{10}$. For the $(L-R)$ model we have $h_6=h_{11}$,
and $h_7=h_{10}$.

Note that the helicity amplitudes describing the contribution of the standard
QED mechanism (see Eq.\,(\ref{eq:Mg4})) and the contribution of the heavy electron (see
Eq.\,(\ref{eq:Mex7})) in the approximation $m=0$ coincide with the results of Ref.
\cite{H86}. Our definition of the helicity amplitudes differs from the
definition of the paper \cite{H86}.

\begin{center}
{\Large{\bf{Appendix C}}}
\end{center}
\setcounter{equation}{0}
\def\theequation{C.\arabic{equation}}

In this Appendix we present the expressions for the polarization observables
in terms of the helicity amplitudes for the case when the initial beams have
arbitrary polarizations.

Let us denote the scattering amplitude from the initial state, where the
electrons and positrons have specific helicities, to the final
$\gamma\gamma $--state by $M_{m\,n}=<\gamma\,\gamma |M|m\,n>,$ where $m\,(n)$ is the
helicity of the positron (electron). In this case the cross section is
$$\frac{d\sigma}{d\Omega}=\frac{1}{256\pi^2\beta s}\bigl\{
\Sigma_0+\xi_z\Sigma_z+\bar\xi_z\bar\Sigma_z+
\xi_z\bar\xi_z\bar\Sigma_{zz}+2\xi_xRe\Sigma_x-
2\xi_yIm\Sigma_x+2\bar\xi_xRe\bar\Sigma_x+2\bar\xi_yIm\bar\Sigma_x+ $$
\begin{equation}\label{C.1}
+2\xi_x\bar\xi_xRe\Sigma_{xx}-2\xi_y\bar\xi_xIm\Sigma_{xx}
+2\xi_x\bar\xi_yIm\bar\Sigma_{xx}+2\xi_y\bar\xi_yRe\bar\Sigma_{xx}+
2\xi_z\bar\xi_xRe\Sigma_{zx}+2\xi_z\bar\xi_yIm\Sigma_{zx}+
\end{equation}
$$+2\xi_x\bar\xi_zRe\bar\Sigma_{zx}-2\xi_y\bar\xi_zIm\bar\Sigma_{zx}
\bigr\}, $$
where
$$\Sigma_0=|M_{+\,+}|^2+|M_{+\,-}|^2+|M_{-\,+}|^2+|M_{-\,-}|^2, \ \
\Sigma_z=|M_{+\,+}|^2+|M_{-\,+}|^2-|M_{+\,-}|^2-|M_{-\,-}|^2, $$
$$\bar\Sigma_z=|M_{-\,-}|^2+|M_{-\,+}|^2-|M_{+\,+}|^2+|M_{+\,-}|^2, \ \
\Sigma_{zz}=|M_{+\,-}|^2+|M_{-\,+}|^2-|M_{+\,+}|^2-|M_{-\,-}|^2, $$
$$\Sigma_x=M_{-\,-}M_{-\,+}^*+M_{+\,-}M_{+\,+}^*, \ \
\bar\Sigma_x=M_{-\,-}M_{+\,-}^*+M_{-\,+}M_{+\,+}^*, $$
$$\Sigma_{xx}=M_{-\,-}M_{+\,+}^*+M_{+\,-}M_{-\,+}^*, \ \
\bar\Sigma_{xx}=M_{-\,-}M_{+\,+}^*-M_{+\,-}M_{-\,+}^*, $$
$$\Sigma_{zx}=M_{-\,+}M_{+\,+}^*-M_{-\,-}M_{+\,-}^*, \ \
\bar\Sigma_{zx}=M_{-\,-}M_{-\,+}^*-M_{+\,-}M_{+\,+}^*, $$
where $\vec {\xi}\,(\vec {\bar\xi})$ is the unit vector in the direction of
the polarization of the electron (positron) in its rest system.

Expressing the $M_{m\,n}$ amplitudes in terms of the helicity amplitudes $H_i$
we obtain the following formulae for the polarization observables
$$\Sigma_0=|H_1|^2+|H_2|^2+|H_3|^2+|H_4|^2+|H_5|^2+|H_6|^2+|H_7|^2+
|H_8|^2+|H_9|^2+|H_{10}|^2+ $$
$$+|H_{11}|^2+|H_{12}|^2+|H_{13}|^2+|H_{14}|^2+|H_{15}|^2+|H_{16}|^2, $$
$$\Sigma_z=|H_1|^2+|H_3|^2+|H_5|^2+|H_7|^2+|H_9|^2+|H_{11}|^2+|H_{13}|^2+
|H_{15}|^2-|H_2|^2-|H_{4}|^2- $$
$$-|H_{6}|^2-|H_{8}|^2-|H_{10}|^2-|H_{12}|^2-|H_{14}|^2-|H_{16}|^2, $$
$$\bar\Sigma_z=|H_3|^2+|H_4|^2+|H_7|^2+|H_8|^2+|H_{11}|^2+|H_{12}|^2+
|H_{15}|^2+|H_{16}|^2-|H_1|^2-|H_{2}|^2- $$
$$-|H_{5}|^2-|H_{6}|^2-|H_{9}|^2-|H_{10}|^2-|H_{13}|^2-|H_{14}|^2, $$
$$\Sigma_{zz}=|H_2|^2+|H_3|^2+|H_6|^2+|H_7|^2+|H_{10}|^2+|H_{11}|^2+
|H_{14}|^2+|H_{15}|^2-|H_1|^2-|H_{4}|^2- $$
$$-|H_{5}|^2-|H_{8}|^2-|H_{9}|^2-|H_{12}|^2-|H_{13}|^2-|H_{16}|^2, $$
$$\Sigma_{x}=H_{2}H_{1}^*+H_{4}H_{3}^*+H_{6}H_{5}^*+H_{8}H_{7}^*+
H_{10}H_{9}^*+H_{12}H_{11}^*+H_{14}H_{13}^*+H_{16}H_{15}^*, $$
$$\bar\Sigma_{x}=H_{3}H_{1}^*+H_{4}H_{2}^*+H_{7}H_{5}^*+H_{8}H_{6}^*+
H_{11}H_{9}^*+H_{12}H_{10}^*+H_{15}H_{13}^*+H_{16}H_{14}^*, $$
$$\Sigma_{xx}=H_{2}H_{3}^*+H_{4}H_{1}^*+H_{6}H_{7}^*+H_{8}H_{5}^*+
H_{10}H_{11}^*+H_{12}H_{9}^*+H_{14}H_{15}^*+H_{16}H_{13}^*, $$
$$\bar\Sigma_{xx}=H_{4}H_{1}^*+H_{8}H_{5}^*+H_{12}H_{9}^*+H_{16}H_{13}^*-
H_{2}H_{3}^*-H_{6}H_{7}^*-H_{10}H_{11}^*-H_{14}H_{15}^*, $$
$$\Sigma_{zx}=H_{3}H_{1}^*+H_{7}H_{5}^*+H_{11}H_{9}^*+H_{15}H_{13}^*-
H_{4}H_{2}^*-H_{8}H_{6}^*-H_{12}H_{10}^*-H_{16}H_{14}^*, $$
$$\bar\Sigma_{zx}=H_{4}H_{3}^*+H_{8}H_{7}^*+H_{12}H_{11}^*+H_{16}H_{15}^*-
H_{2}H_{1}^*-H_{6}H_{5}^*-H_{10}H_{9}^*-H_{14}H_{13}^*. $$

The differential cross section of the $e^+e^-\to \gamma\gamma $ reaction
in the case when the initial beams have arbitrary polarizations can be
presented in the following general form (valid in the approximation of the
real amplitudes)
\begin{equation}\label{C.2}
\frac{d\sigma}{d\Omega}=\frac{d\sigma_0}{d\Omega}
\bigl\{1+\xi_zA_z+\bar\xi_z\bar A_z+\xi_xA_x+\bar\xi_x\bar A_x+
\xi_x\bar\xi_x\bar A_{xx}+\xi_y\bar\xi_y\bar A_{yy}+
\end{equation}
$$\xi_z\bar\xi_z\bar A_{zz}+\xi_z\bar\xi_x\bar A_{zx}+
\xi_x\bar\xi_z\bar A_{xz}\bigr\},  $$
where $d\sigma_0/d\Omega $ is the differential cross section of the
$e^+\,e^-\to \gamma\,\gamma $ reaction for the case when all particles unpolarized.

Let us consider the contribution of the mechanisms under consideration
(pure QED, contact $e\,e\,\gamma\,\gamma $ interaction and heavy lepton excitation)
to the asymmeties $A_i$ and $A_{i\,j}$

1. \underline{The contribution of the contact $e\,e\,\gamma\,\gamma $ interaction}.
In this case the asymmetries and cross secions are
\begin{equation}\label{C.3}
\frac{d\sigma_0}{d\Omega}=\frac{\alpha^2}{\beta s}H,
\end{equation}
$$H=(1-\beta^2x^2)^{-2}[\beta^2(1-x^4)+d^2(1+\beta^2z)]+
\frac{a}{4}R\beta^2(1-\beta^2x^2)^{-1}(1-x^4+d^2z)+ $$
$$+\frac{R^2}{64}[\beta^2(1-x^4)(a^2+\beta^2b^2)+d^2(b^2+\beta^2a^2z)], $$
$$A_z=-\frac{b}{4}\beta RH^{-1}[\beta^2(1-x^4)-d^2][(1-\beta^2x^2)^{-1}
+\frac{a}{8}R], $$
$$\bar A_z=-\frac{b}{4}\beta RH^{-1}[\beta^2(1-x^4)+d^2][(1-\beta^2x^2)^{-1}
+\frac{a}{8}R], $$
$$A_x=-\frac{b}{4}\beta^3 RH^{-1}dx(1-x^2)^{3/2}[(1-\beta^2x^2)^{-1}
+\frac{a}{8}R], $$
$$\bar A_x=A_x, $$
\begin{equation}\label{C.4}
A_{zz}=H^{-1}\bigl\{(1-\beta^2x^2)^{-2}[\beta^2(1-x^4)-d^2(1+\beta^2z)]+
\frac{a}{4}R\beta^2(1-\beta^2x^2)^{-1}(1-x^4-d^2z)+
\end{equation}
$$+\frac{R^2}{64}[\beta^2(1-x^4)(a^2+\beta^2b^2)-d^2(b^2+\beta^2a^2z)], $$
$$A_{xx}=H^{-1}\bigl\{(1-\beta^2x^2)^{-2}[\beta^2(1-x^2)^2+d^2(\beta^2z-1)]+
\frac{a}{4}R\beta^2(1-\beta^2x^2)^{-1}[d^2z+(1-x^2)^2]+ $$
$$+\frac{R^2}{64}[d^2(\beta^2a^2z-b^2)+\beta^2(1-x^2)^2(a^2-
\beta^2b^2)]\bigr\}, $$
$$A_{yy}=H^{-1}\bigl\{(1-\beta^2x^2)^{-2}[-\beta^2(1-x^2)^2+d^2(\beta^2z-1)]+
\frac{a}{4}R\beta^2(1-\beta^2x^2)^{-1}[d^2z-(1-x^2)^2]+ $$
$$+\frac{R^2}{64}[d^2(\beta^2a^2z-b^2)-\beta^2(1-x^2)^2(a^2-
\beta^2b^2)]\bigr\}, $$
$$A_{zx}=2H^{-1}\beta^2dx(1-x^2)^{3/2}\bigl [(1-\beta^2x^2)^{-2}
+\frac{a}{4}R(1-\beta^2x^2)^{-1}+\frac{R^2}{64}a^2\bigr ], $$
$$A_{xz}=A_{zx}, \ \ d=\frac{m}{E}, $$
where $z=1+(1-x^2)^2, R=s^2/\Lambda^4.$ Note that at $R=0$ these formulae
describe the contribution of the pure QED mechanism, terms proportional to
$R$ describe the contribution of the interference QED mechanism and
contact $ee\gamma\gamma $ interaction, and the terms $\sim R^2$ -
the contribution of the square of the contact interaction.

Note that the asymmetries $A_x,\, \bar A_x,\, A_z,\, \bar A_z$ are not zero is due
to the presence of the space--parity violating term in the contact
$e\,e\gamma\,\gamma $ interaction, which is described by the parameter $b$.

The polarization observables for the $e^+\,e^-\to \gamma\,\gamma $ reaction (both
initial beams are polarized) for the case of the pure QED mechanism is
$$A_{xx}=H_k^{-1}(1-\beta^2x^2)^{-2}[\beta^2(1-x^2)^2+d^2(\beta^2z-1)], $$
$$A_{yy}=H_k^{-1}(1-\beta^2x^2)^{-2}[-\beta^2(1-x^2)^2+d^2(\beta^2z-1)], $$
$$A_{zz}=H_k^{-1}(1-\beta^2x^2)^{-2}[\beta^2(1-x^4)-d^2(\beta^2z+1)], $$
$$A_{zx}=2H_k^{-1}\beta^2dx(1-x^2)^{3/2}(1-\beta^2x^2)^{-2}, \ \
A_z=\bar A_z=A_x=\bar A_x=0, $$
$$H_k=(1-\beta^2x^2)^{-2}[\beta^2(1-x^4)+d^2(\beta^2z+1)]. $$
At high energies one can neglect by the electron mass and in this case the
expressions for the polarization observables are essentially simplified.
They are
\begin{equation}\label{C.5}
\frac{d\sigma_0}{d\Omega}=\frac{\alpha^2}{s}H_0,
\end{equation}
$$H_0=(1+x^2)(1-x^2)^{-1}+\frac{a}{4}R(1+x^2)+\frac{R^2}{64}
(1-x^4)(a^2+b^2), $$
$$A_z=-2bR(1-x^2)[8+aR(1-x^2)][64+16aR(1-x^2)+
R^2(1-x^2)^2(a^2+b^2)]^{-1}, $$
$$\bar A_z=A_z, \ \ A_{zz}=1, \ \ A_x=\bar A_x=A_{zx}=A_{xz}=0, $$
$$A_{xx}=H_0^{-1}[1+\frac{a}{4}R(1-x^2)+\frac{R^2}{64}
(1-x^2)^2(a^2-b^2),  \ \  A_{yy}=-A_{xx}. $$
The pure QED mechanism in this approximation is described by
$$\frac{d\sigma_0}{d\Omega}=\frac{\alpha^2}{s}\frac{1+x^2}{1-x^2}, $$
$$A_{zz}=1, \ \ A_{xx}=-A_{yy}=\frac{1+x^2}{1-x^2}, \ \
A_z=\bar A_z=A_x=\bar A_x=A_{zx}=A_{xz}=0. $$
2. \underline{The contribution of the heavy lepton}. In this case the
asymmetries and cross secions are
\begin{equation}\label{C.6}
\frac{d\sigma_0}{d\Omega}=\frac{\alpha^2}{\beta s}I,
\end{equation}
$$I=(1-\beta^2x^2)^{-2}[\beta^2(1-x^4)+d^2(1+\beta^2z)]-2\frac{\lambda^2}{y}
(1-\beta^2x^2)^{-1}(Q^2-\beta^2x^2)^{-1} $$
$$\bigl\{\beta^2(1-x^2)[2\beta^2x^2
+Q(1+x^2)]+Qd^2[d\sqrt{2y}+\beta^2(1-x^2)^2]\bigr\}+ $$
$$+\frac{\lambda^4}{y^2}(Q^2-\beta^2x^2)^{-2}\bigl\{4Q\beta^2x^2[2y+
\beta^2(1-x^2)]+Q^2[2y(1+\beta^2)+\beta^2(1-x^2)(1+d^2+x^2)]+ $$
$$+\beta^2x^2[2y(d^2+2\beta^2x^2)+\beta^4(1-x^4)]\bigr\}, $$
$$IA_{zz}=(1-\beta^2x^2)^{-2}[\beta^2(1-x^4)-d^2(1+\beta^2z)]-2\frac{\lambda^2}{y}
(1-\beta^2x^2)^{-1}(Q^2-\beta^2x^2)^{-1} $$
$$\bigl\{\beta^2(1-x^2)[2\beta^2x^2
+Q(1+x^2)]-Qd^2[d\sqrt{2y}+\beta^2(1-x^2)^2]\bigr\}+ $$
$$+\frac{\lambda^4}{y^2}(Q^2-\beta^2x^2)^{-2}\bigl\{4Q\beta^2x^2[-2y+
\beta^2(1-x^2)]+Q^2[-2y(1+\beta^2)+\beta^2(1-x^2)(\beta^2+d^2x^2+x^2)]+ $$
$$+\beta^2x^2[2y(d^2-2x^2)+\beta^4(1-x^4)]\bigr\}, $$
$$IA_{xx}=(1-\beta^2x^2)^{-2}[\beta^2(1-x^2)^2+d^2(\beta^2z-1)]-2\frac{\lambda^2}{y}
(1-\beta^2x^2)^{-1}(Q^2-\beta^2x^2)^{-1} $$
$$\bigl\{Q\beta^2(1+d^2)(1-x^2)^2
-d\sqrt{2y}[2\beta^2x^2+Q(1+\beta^2)]\bigr\}+ $$
$$+\frac{\lambda^4}{y^2}(Q^2-\beta^2x^2)^{-2}\bigl\{Q^2[-2yd^2+
\beta^2(1+d^2)(1-x^2)^2]+\beta^2x^2[2yd^2(2x^2-1)-\beta^4(1-x^2)^2]
\bigr\}, $$
$$IA_{yy}=(1-\beta^2x^2)^{-2}[-\beta^2(1-x^2)^2+d^2(\beta^2z-1)]+
2\frac{\lambda^2}{y}(1-\beta^2x^2)^{-1}(Q^2-\beta^2x^2)^{-1} $$
$$\bigl\{Q\beta^4(1-x^2)^2+d\sqrt{2y}[2\beta^2x^2+Q(1+\beta^2)]\bigr\}-
\frac{\lambda^4}{y^2}(Q^2-\beta^2x^2)^{-1}[2yd^2+\beta^4(1-x^2)^2], $$
$$IA_{zx}=2\beta^2dx(1-x^2)^{1/2}\bigl\{(1-\beta^2x^2)^{-2}(1-x^2)-
\frac{\lambda^2}{y}(1-\beta^2x^2)^{-1}(Q^2-\beta^2x^2)^{-1} $$
$$[d\sqrt{2y}+(1-x^2)(2Q+\beta^2)]+
\frac{\lambda^4}{y^2}(Q^2-\beta^2x^2)^{-2}[Q(Q+\beta^2)(1-x^2)
-2y(Q^2+x^2)]\bigr\}, $$
$$A_{xz}=A_{zx},   \ \ A_z=\bar A_z=A_x=\bar A_x=0. $$
where $y=2M^2/s, \ Q=d^2/2-1-y, $ $M$ is the heavy lepton mass. Note that
at $\lambda=0$ these formulae describe the observables caused by the
contribution of the QED mechanism for this reaction, the terms proportional
to $\lambda^2$ describe the contribution of the interference of the QED
mechanism and heavy lepton excitation and the terms proportional to $\lambda^4$
 - the contribution of the square of the heavy lepton excitation term.

At high energies one can neglect by the electron mass and we have
$$\frac{d\sigma_0}{d\Omega}=\frac{\alpha^2}{s}I_0, $$
$$I_0=(1+x^2)(1-x^2)^{-1}+2\frac{\lambda^2}{y}[(1+y)^2-x^2]^{-1}
[1-x^2+y(1+x^2)]+  $$
$$+\frac{\lambda^4}{y^2}[(1+y^2)^{2}-x^2]^{-2}
[4y^3+y^2(1-x^2)(x^2+9)+6y(1-x^2)^2+(1-x^2)^3], $$
$$I_0A_{zz}=(1+x^2)(1-x^2)^{-1}+2\frac{\lambda^2}{y}[(1+y)^2-x^2]^{-1}
[1-x^2+y(1+x^2)]+  $$
$$+\frac{\lambda^4}{y^2}[(1+y^2)^2-x^2]^{-2}
[-4y^3+y^2(1-x^2)(x^2-7)-2y(1-x^2)^2+(1-x^2)^3], $$
$$I_0A_{xx}=1+\frac{\lambda^2}{y}(1-x^2)[(1+y)^2-x^2]^{-1}
[2(1+y)+\frac{\lambda^2}{y}(1-x^2)], $$
$$A_{yy}=-A_{xx},   \ \ A_z=\bar A_z=A_x=\bar A_x=A_{zx}=A_{xz}=0. $$

\begin{center}
{\Large{\bf{Appendix D}}}
\end{center}
\setcounter{equation}{0}
\def\theequation{D.\arabic{equation}}

In this Appendix we present the expressions for the polarization observables
in terms of the helicity amplitudes for the case when both photons have
arbitrary polarizations which is described by the Stocks parameters and
initial beams are unpolarized.

We use the following four$-$dimensional representation \cite{BLP} for the
spin--density matrix of the partially polarized photon
\begin{equation}\label{D.1}
\rho_{\mu\nu}=\frac{1}{2}(e_{\mu}^{(1)}e_{\nu}^{(1)}+
e_{\mu}^{(2)}e_{\nu}^{(2)})+
\frac{\eta_1}{2}(e_{\mu}^{(1)}e_{\nu}^{(2)}+
e_{\mu}^{(2)}e_{\nu}^{(1)})-
\end{equation}
$$\frac{i}{2}\eta_2(e_{\mu}^{(1)}e_{\nu}^{(2)}-
e_{\mu}^{(2)}e_{\nu}^{(1)})+
\frac{\eta_3}{2}(e_{\mu}^{(1)}e_{\nu}^{(1)}-
e_{\mu}^{(2)}e_{\nu}^{(2)}), $$
where $\eta_1, \ \eta_2, \ \eta_3 $ are the real Stocks parameters. The
space-like unit real four--vectors $e_{\mu}^{(1)}$ and $e_{\mu}^{(2)}$ are
orthogonal each other and photon four--momentum $k$:
$$e^{(1)^2}=e^{(2)^2}=-1, \ \ e^{(1)}\times e^{(2)}=0, \ \
e^{(1)}\times k=e^{(2)}\times k=0. $$
Let us consider the following four-vectors constructed with the help of
the four--momenta of particles in the $e^+e^-\to \gamma\gamma $ reaction
$$P_{\mu}=p_{1\mu}-p_{2\mu}-K_{\mu}K^{-2}(p_1\times K-p_2\times K), \ \
N_{\mu}=\epsilon_{\mu\nu\rho\sigma}P_{\nu}q_{\rho]}K_{\sigma}, $$
where $K=k_1-k_2, \ q=p_1+p_2=k_1+k_2.$ They are evidently
orthogonal each other.They are also orthogonal to the vectors $k_1$
and $k_2$. These vectros are space--like ones since they are
orthogonal to the time--like four--vector $q (q^2>0)$. Normalize
them to unit magnitude
$$e_{\mu}^{(1)}=\frac{N_{\mu}}{\sqrt{-N^2}}, \ \
e_{\mu}^{(2)}=\frac{P_{\mu}}{\sqrt{-P^2}}, $$
we obtain a couple of four--vectors having all required properties. For each
of two photons these four--vectors can serve as four--dimensional orths and
with the help of these vectors we can construct their spin--density matrices.

Let us denote the scattering amplitude from the initial unpolarized state
of the $e^+e^-$--pair to the final $\gamma\gamma $--state, where both photons
have definite helicity, by $L_{m\,n}=<m\,n|M|e^+\,e^->,$ where $m(n)$ is the
helicity of the first (second) photon. The photon spin--density matrix will
be described by the standard Stocks parameters. In this case the cross section
can be written in the following general form
$$\frac{d\sigma}{d\Omega}=\frac{1}{1024\pi^2\beta s}\bigl\{
\Omega_0+\eta_2\Omega_2+\bar\eta_2\bar\Omega_2+
\eta_2\bar\eta_2\bar\Omega_{22}+2\eta_3Re\Omega_3-
2\eta_1Im\Omega_3+2\bar\eta_3Re\bar\Omega_3+2\bar\eta_1Im\bar\Omega_3+ $$
$$+2\eta_3\bar\eta_3Re\Omega_{33}+2\eta_1\bar\eta_3Im\Omega_{33}
-2\eta_1\bar\eta_1Re\Omega_{11}-2\eta_3\bar\eta_1Im\Omega_{11}+
2\eta_2\bar\eta_3Re\Omega_{23}-2\eta_2\bar\eta_1Im\Omega_{23}- $$
\begin{equation}\label{D.2}
-2\eta_3\bar\eta_2Re\Omega_{32}-2\eta_1\bar\eta_2Im\Omega_{32}
\bigr\},
\end{equation}
where we have
$$\Omega_0=|L_{++}|^2+|L_{+-}|^2+|L_{-+}|^2+|L_{--}|^2, \ \
\Omega_2=-|L_{++}|^2+|L_{-+}|^2-|L_{+-}|^2+|L_{--}|^2, $$
$$\bar\Omega_2=-|L_{--}|^2-|L_{-+}|^2+|L_{++}|^2-|L_{+-}|^2, \ \
\Omega_{22}=|L_{+-}|^2+|L_{-+}|^2-|L_{++}|^2-|L_{--}|^2, $$
$$\Omega_3=L_{++}L_{-+}^*+L_{+-}L_{--}^*, \ \
\bar\Omega_3=L_{++}L_{+-}^*+L_{-+}L_{--}^*, $$
$$\Omega_{33}=L_{--}L_{++}^*+L_{-+}L_{+-}^*, \ \
\Omega_{11}=L_{+-}L_{-+}^*-L_{++}L_{--}^*, $$
$$\Omega_{23}=L_{--}L_{-+}^*-L_{+-}L_{++}^*, \ \
\Omega_{32}=L_{--}L_{+-}^*-L_{-+}L_{++}^*, $$
and $\eta_i \,(\bar\eta_i)$ (where $i=1,\,2,\,3$) are the Stocks parameters of the
first (second) photon.

Expressing the $L_{mn}$ amplitudes in terms of the helicity amplitudes $H_i$
we obtain the following formulae for the polarization observables
$$\Omega_0=\Sigma_0, $$
$$\Omega_2=-|H_1|^2-|H_2|^2-|H_3|^2-|H_4|^2-|H_5|^2-|H_6|^2-|H_7|^2-
|H_8|^2+|H_9|^2+|H_{10}|^2+ $$
$$+|H_{11}|^2+|H_{12}|^2+|H_{13}|^2+|H_{14}|^2+|H_{15}|^2+|H_{16}|^2, $$
$$\bar\Omega_2=|H_1|^2+|H_2|^2+|H_3|^2+|H_4|^2+|H_9|^2+|H_{10}|^2+|H_{11}|^2+
|H_{12}|^2-|H_5|^2-|H_{6}|^2- $$
$$-|H_{7}|^2-|H_{8}|^2-|H_{13}|^2-|H_{14}|^2-|H_{15}|^2-|H_{16}|^2, $$
$$\Omega_{22}=-|H_1|^2-|H_2|^2-|H_3|^2-|H_4|^2-|H_{13}|^2-|H_{14}|^2-
|H_{15}|^2-|H_{16}|^2+|H_5|^2+|H_{6}|^2+ $$
$$+|H_{7}|^2+|H_{8}|^2+|H_{9}|^2+|H_{10}|^2+|H_{11}|^2+|H_{12}|^2, $$
$$\Omega_{3}=H_{1}H_{9}^*+H_{2}H_{10}^*+H_{3}H_{11}^*+H_{4}H_{12}^*+
H_{5}H_{13}^*+H_{6}H_{14}^*+H_{7}H_{15}^*+H_{8}H_{16}^*, $$
$$\bar\Omega_{3}=H_{1}H_{5}^*+H_{2}H_{6}^*+H_{3}H_{7}^*+H_{4}H_{8}^*+
H_{9}H_{13}^*+H_{10}H_{14}^*+H_{11}H_{15}^*+H_{12}H_{16}^*, $$
$$\Omega_{33}=H_{13}H_{1}^*+H_{14}H_{2}^*+H_{15}H_{3}^*+H_{16}H_{4}^*+
H_{9}H_{5}^*+H_{10}H_{6}^*+H_{11}H_{7}^*+H_{12}H_{8}^*, $$
$$\Omega_{11}=H_{5}H_{9}^*+H_{6}H_{10}^*+H_{7}H_{11}^*+H_{8}H_{12}^*-
H_{1}H_{13}^*-H_{2}H_{14}^*-H_{3}H_{15}^*-H_{4}H_{16}^*, $$
$$\Omega_{23}=H_{13}H_{9}^*+H_{14}H_{10}^*+H_{15}H_{11}^*+H_{16}H_{12}^*-
H_{5}H_{1}^*-H_{6}H_{2}^*-H_{7}H_{3}^*-H_{8}H_{4}^*, $$
$$\Omega_{32}=H_{13}H_{5}^*+H_{14}H_{6}^*+H_{15}H_{7}^*+H_{16}H_{8}^*-
H_{9}H_{1}^*-H_{10}H_{2}^*-H_{11}H_{3}^*-H_{12}H_{4}^*. $$

The differential cross section of the $e^+\,e^-\to \gamma\,\gamma $ reaction
in the case when the initial beams are unpolarized and final photons have
arbitrary polarizations can be presented in the following general form
(valid in the approximation of the real amplitudes)
\begin{equation}\label{D.3}
\frac{d\sigma}{d\Omega}=\frac{1}{4}\frac{d\sigma_0}{d\Omega}
\bigl\{1+\eta_2C_2+\bar\eta_2\bar C_2+\eta_3C_3+\bar\eta_3\bar C_3+
\eta_2\bar\eta_2 C_{22}+\eta_3\bar\eta_3C_{33}+
\eta_1\bar\eta_1C_{11}+\eta_2\bar\eta_3C_{23}+
\eta_3\bar\eta_2C_{32}\bigr\}.  
\end{equation}
Let us consider the contribution of the mechanisms under consideration
(pure QED, contact $e\,e\,\gamma\,\gamma $ interaction and heavy lepton excitation)
to the photon polarization observables

1. \underline{The contribution of the contact $ee\gamma\gamma $ interaction}.
In this case the photon polarization observables are described by
$$C_2=\frac{b}{4}RH^{-1}\bigl\{(1-\beta^2x^2)^{-1}[d^2-2\beta^3x(1-x^2)]-
\frac{R}{4}\beta^3ax(1-x^2)\bigr\}, $$
$$\bar C_2=-\frac{b}{4}RH^{-1}\bigl\{(1-\beta^2x^2)^{-1}[d^2+2\beta^3x(1-x^2)]
+\frac{R}{4}\beta^3ax(1-x^2)\bigr\}, $$
$$C_{22}=-H^{-1}\bigl\{(1-\beta^2x^2)^{-2}[\beta^2(\beta^2x^4-1)+
d^2(1+2\beta^2x^2)]-\frac{a}{4}R\beta^2(1-\beta^2x^2)^{-1}(1-
2d^2x^2-\beta^2x^4)+ $$
$$+\frac{1}{64}R^2[d^2(\beta^2a^2+b^2)-d^2\beta^2a^2
(1-x^2)^2-\beta^2(a^2+\beta^2b^2)(1-x^4)]\bigr\}, $$
\begin{equation}\label{D.4}
C_{3}=-2H^{-1}\beta^2d^2(1-x^2)\bigl [(1-\beta^2x^2)^{-2}
+\frac{a}{4}R(1-\beta^2x^2)^{-1}+\frac{1}{64}a^2R^2\bigr ], 
\end{equation}
$$\bar C_3=C_3, $$
$$C_{33}=H^{-1}\bigl\{-(1-\beta^2x^2)^{-2}[\beta^4(1-x^2)^2+d^4]-
\frac{a}{4}R\beta^2(1-\beta^2x^2)^{-1}[(1-x^2)^2-d^2z]+ $$
$$+
\frac{1}{64}R^2[d^2(\beta^2a^2z+b^2)-\beta^2(a^2+
\beta^2b^2)(1-x^2)^2]\bigr\}, $$
$$C_{11}=-H^{-1}\bigl\{(1-\beta^2x^2)^{-2}[d^4-\beta^4(1-x^2)^2]-
\frac{a}{4}R\beta^2(1-\beta^2x^2)^{-1}[\beta^2(1-x^2)^2+d^2]+ $$
$$+\frac{1}{64}R^2[-d^2(\beta^2a^2+b^2)-\beta^4(a^2+
b^2)(1-x^2)^2]\bigr\}, $$
$$C_{23}=C_{32}=0. $$

Note that at $R=0$ these formulae describe the contribution of the pure QED
mechanism, terms proportional to $R$ describe the contribution of the
interference QED mechanism and contact $e\,e\,\gamma\,\gamma $ interaction, and the
terms $\sim R^2$ - the contribution of the square of the contact interaction.

Note that the observables $C_2,\, \bar C_2$ are not zero due
to the presence of the space-parity violating term in the contact
$e\,e\,\gamma\,\gamma $ interaction, which is described by the parameter $b$.

The photon polarization observables for the $e^+\,e^-\to \gamma\,\gamma $ reaction
(both final photons are polarized) for the case of the pure QED mechanism are
\begin{equation}\label{D.5}
C_{22}=-H_k^{-1}(1-\beta^2x^2)^{-2}[\beta^2(\beta^2x^4-1)^2+
d^2(1+2\beta^2x^2)],
\end{equation}
$$C_{3}=-2H_k^{-1}(1-\beta^2x^2)^{-2}\beta^2d^2(1-x^2), \ \ \bar C_3=C_3, $$
$$C_{33}=-H_k^{-1}(1-\beta^2x^2)^{-2}[\beta^4(1-x^2)^2+d^4], $$
$$C_{11}=-H_k^{-1}(1-\beta^2x^2)^{-2}[d^4-\beta^4(1-x^2)^2], \ \
C_2=\bar C_2=C_{23}=C_{32}=0. $$

At high energies one can neglect by the electron mass and in this case the
expressions for the photon polarization observables are essentially
simplified. They are
$$\frac{d\sigma_0}{d\Omega}=\frac{\alpha^2}{4s}H_0, $$
\begin{equation}\label{D.6}
C_2=-\frac{1}{2}b\,R\,x\,H_0^{-1}[1+\frac{a}{8}R(1-x^2)], \ \ 
\bar C_2=-C_2, \ \  C_3=\bar C_3=C_{23}=C_{32}=0, 
\end{equation}
$$C_{22}=GH_0^{-1}\frac{1+x^2}{1-x^2}, \ \ C_{33}=-GH_0^{-1}, \ \
C_{11}=-C_{33}, \ \ G=1+\frac{a}{4}R(1-x^2)+
\frac{R^2}{64}(1-x^2)^2(a^2+b^2).   $$
The pure QED mechanism in this approximation is described by
$$\frac{d\sigma_0}{d\Omega}=\frac{\alpha^2}{s}\frac{1+x^2}{1-x^2}, $$
$$C_{22}=1, \ \ C_{33}=-C_{11}=-\frac{1+x^2}{1-x^2}, \ \
C_2=\bar C_2=C_3=\bar C_3=C_{23}=C_{32}=0. $$

2. \underline{The contribution of the heavy lepton}. In this case the
the photon polarization observables are described by
$$IC_{22}=(1-\beta^2x^2)^{-2}[1+\beta^2(\beta^2x^4+2d^2x^2-2)]-
2\frac{\lambda^2}{y}(1-\beta^2x^2)^{-1}(Q^2-\beta^2x^2)^{-1} $$
$$\bigl\{Q[d^3\sqrt{2y}+\beta^2(\beta^2x^4+2d^2x^2-1-d^2)]-
2\beta^4x^2(1-x^2)\bigr\}+ $$
$$+\frac{\lambda^4}{y^2}(Q^2-\beta^2x^2)^{-2}\bigl\{q^2[2\beta^4x^4+
\beta^2x^2(d^2+4Q)+Q^2(1+\beta^2)]-$$
$$\beta^2(1-x^2)[(1+d^2)Q^2+
\beta^2x^2(Q^2+4Q+\beta^2+\beta^2x^2)], $$
$$IC_{3}=2\beta^2d(1-x^2)\bigl\{d(1-\beta^2x^2)^{-2}
+\frac{\lambda^2}{y}(Q^2-\beta^2x^2)^{-1}[Q(\sqrt{2y}-d)(1-\beta^2x^2)^{-1}
-\frac{\lambda^2}{y}-\sqrt{2y}]\bigr\}, $$
\begin{equation}\label{D.7}
IC_{33}=-(1-\beta^2x^2)^{-2}[\beta^4(1-x^2)^2+d^4]+
2\frac{\lambda^2}{y}(1-\beta^2x^2)^{-1}(Q^2-\beta^2x^2)^{-1}\times 
\end{equation}
$$\bigl\{Q\beta^4(1-x^2)^2+d\sqrt{2y}[2\beta^2x^2+Q(1+\beta^2)]\bigr\}-
\frac{\lambda^4}{y^2}(Q^2-\beta^2x^2)^{-1}[2yd^2+\beta^4(1-x^2)^2], $$
$$IC_{11}=(1-\beta^2x^2)^{-2}[\beta^4(1-x^2)^2-d^4]+
2\frac{\lambda^2}{y}(1-\beta^2x^2)^{-1}(Q^2-\beta^2x^2)^{-1}\times $$
$$\bigl\{d\sqrt{2y}[Q(1+\beta^2)+2\beta^2x^2]-Q\beta^4(1-x^2)^2\bigr\}+
\frac{\lambda^4}{y^2}(Q^2-\beta^2x^2)^{-1}[\beta^4(1-x^2)^2-2yd^2], $$
$$\bar C_{3}=C_{3},   \ \ C_2=\bar C_2=C_{32}=C_{23}=0. $$

Note that at $\lambda=0$ these formulae describe the photon polarization
observables caused by the contribution of the QED mechanism for this reaction,
the terms proportional to $\lambda^2$ describe the contribution of the
interference of the QED mechanism and heavy lepton excitation and the terms
proportional to $\lambda^4$  - the contribution of the square of the heavy
lepton excitation term.

At high energies one can neglect by the electron mass and we have
$$I_0C_{22}=\frac{1+x^2}{1-x^2}+2\frac{\lambda^2}{y}[(1+y)^2-x^2]^{-1}
[1-x^2+y(1+x^2)]+  $$
$$+\frac{\lambda^4}{y^2}[(1+y^2)^2-x^2]^{-2}
[-4y^3+y^2(1-x^2)(x^2-7)-2y(1-x^2)^2+(1-x^2)^3], $$
$$I_0C_{11}=1+\frac{\lambda^2}{y}(1-x^2)[(1+y)^2-x^2]^{-1}
[2(1+y)+\frac{\lambda^2}{y}(1-x^2)], $$
$$C_{33}=-C_{11},   \ \ C_3=\bar C_3=C_2=\bar C_2=C_{32}=C_{23}=0. $$

\end{document}